\documentclass[prb,aps,amsmath,amssymb,reprint,showpacs,floatfix,superscriptaddress,longbibliography]{revtex4-2}
\usepackage{graphicx}% Include figure files
\usepackage{dcolumn}% Align table columns on decimal point
\usepackage{bm}% bold math
\usepackage{mathtools}% math tools
\usepackage{placeins}% float management
\usepackage{enumitem}
\usepackage{xcolor}

\begin{document}

%\preprint{Test}

\title{
Electrostatic gating and the
interference of chiral Majoranas
in thin slabs of magnetic topological insulators
} %Title of paper



\author{Javier Osca}
\email{javier.osca@uib.cat}
\affiliation{Institute for Cross-Disciplinary Physics and Complex Systems IFISC (CSIC-UIB), E-07122 Palma, Spain} 
\affiliation{Department of Physics, University of the Balearic Islands, E-07122 Palma, Spain}
\author{Llorenç Serra}
\email{llorens.serra@uib.es}
\affiliation{Institute for Cross-Disciplinary Physics and Complex Systems IFISC (CSIC-UIB), E-07122 Palma, Spain} 
\affiliation{Department of Physics, University of the Balearic Islands, E-07122 Palma, Spain}

\date{July 21, 2025}

\begin{abstract}
We study the interference of chiral Majoranas 
in a magnetic topological insulator thin slab having a grounded section proximity coupled to a superconductor and another section under the influence of top-bottom electrostatic gating.
The gated section locally widens an energy gap and  
mediates the coupling between the quantum anomalous Hall states of the leads and the chiral Majorana states of the proximitized sector. 
Local and non-local conductances offer measurable hints of the existence of transport mediated by chiral Majorana modes.
Local conductances on the two leads reveal characteristic oscillatory patterns as a function of the gating strength, with peculiar correlations 
depending on the distance between gated and proximitized sectors.
A gate tunable Majorana diode effect on nonlocal conductances emerges when the chemical 
potential deviates from zero. 
We suggest a protocol to identify chiral Majorana physics based on 
a sequence of electrostatic gates that allows the tuning of 
chiral Majorana interference. 
\end{abstract}

%\pacs{73.63.Nm,74.45.+c}% insert suggested PACS numbers in braces on next line

\maketitle %\maketitle must follow title, authors, abstract and \pacs
%\tableofcontents

\section{Introduction}

In condensed matter systems, Majorana modes
are zero-energy quasiparticle excitations that are their own antiparticles (see, e.g., Refs.\ \cite{Ali2012,Bee2013,Elliot2015,Sato2016}).
They may appear
as lower-dimensional modes in systems with particle-hole symmetry,
such as
topological superconductors \cite{Schny2008,Kita2009,Qi2011,Sato2017}.
In particular, two types of Majorana modes have 
attracted strong interest:
a) Majorana bound states at the ends of nearly 1D 
superconducting systems, such as those based on semiconductors \cite{Prada2020,Laub2021},
magnetic topological insulators (MTI) \cite{Zeng18,Chen18,Atan2023} or 
atomic and quantum dot chains \cite{Nadj2014,Sau2012};  
b) Majorana chiral modes that emerge as propagating states in the 1D edges of 2D 
MTI's,
when the necessary symmetries and conditions are met \cite{Qi2010}.  
Chiral Majoranas are propagating states
analogous to quantum anomalous Hall (QAH) and quantum spin Hall (QSH) edge states but in superconducting materials \cite{Kon2008,Qi2011,Chang2013,Qiu2022}.

Both Majorana bound states and chiral modes are promising candidates for topological quantum computation \cite{Nay08} but their detection remains challenging. 
The same robustness that characterises these states also makes their detection arduous \cite{Kay2020,Agh2023,Uday2024,Uday2025,Agh2025}. 
In particular, electrical measurements of Majorana bound states cannot
easily be distinguished from those of bound Andreev levels,
which may be induced by material disorder. 
An important aspect of chiral Majoranas, as opposed to Majorana bound states,
is that they connect the different leads of a device via quasiparticle currents.
These edge chiral currents offer a high degree of protection against backscattering. 
Therefore, it should be possible
to identify distinctive interference correlations in charge transport or other quasiparticle properties
in the presence of Majorana chiral modes. 
This would provide 
unambiguous evidence of
the existence of such topological states. 

\begin{figure}[t]
  \centering
  \includegraphics[width=0.5\textwidth]{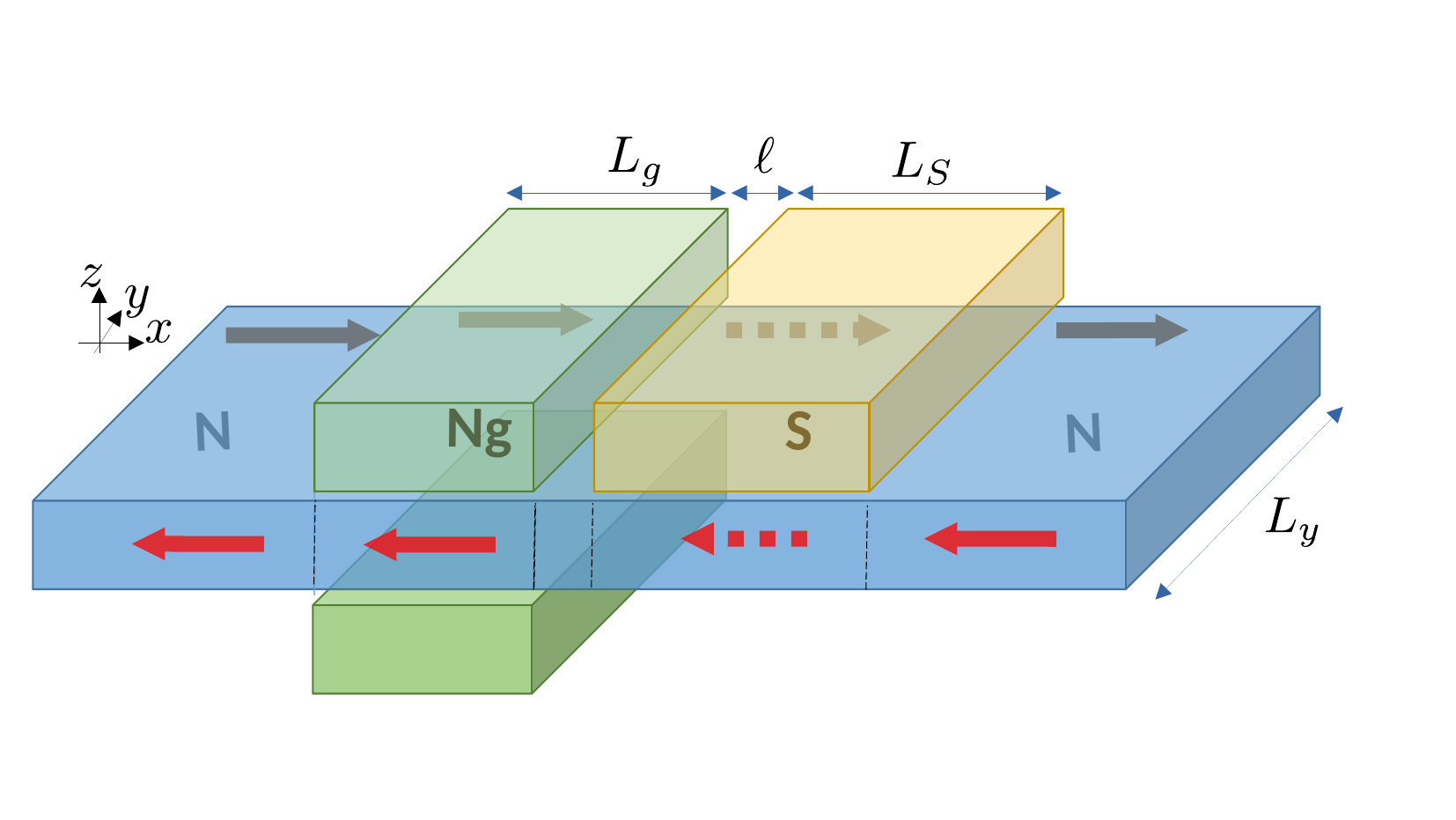}
  \caption{Schematic of an MTI thin slab (blue) with
a proximitized superconducting sector (S, superconductor in yellow) and a normal gated section (Ng, top/bottom electrodes in green). 
The arrows qualitatively indicate transport by chiral edge 
Fermionic (solid) and Majorana (dashed)  modes in different sectors of the slab. The figure also shows the definitions of the transverse width ($L_y$)
and the lengths of the 
proximitized ($L_S$), gated ($L_g$) and separation ($\ell$) distances
along the slab.  
  }
  \label{F0}
\end{figure}

MTI's are 3D topological insulators with ferromagnetic ordering \cite{Toku2019,Chang2023}. It is theorized that, when in contact with a metallic s-wave superconductor, they can exhibit topological phases  becoming 
topological superconductors in which chiral Majoranas may appear as edge states on the boundaries of MTI thin films \cite{Fu2008,Qi2010,Wang2015}.

In this paper, we focus on chiral Majorana physics in QAH slabs
of MTI. 
We consider a relatively narrow QAH slab part of which is in contact with a grounded s-wave superconductor, as in 
Refs.\ \cite{Wang2015,Osca2018}. We also consider an additional section of the 
slab under the influence of a tunable 
perpendicular electric field (see Fig.\ \ref{F0}).
Cases of uniform gating throughout the slab have been discussed in 
Refs.\ \cite{Wang2015b,Wang2016,Chong2023}.
The presence of a gated sector breaks the left-right inversion symmetry of the device, affecting  the local and nonlocal conductances 
in the terminals 
in characteristic ways
depending on the topological phase
of the superconductor.  
In this paper we refer to regions carrying the QAH state as
'normal'
and regions 
holding chiral Majorana edge states as
'superconductor'.

By tuning the gate, correlated and anticorrelated local conductance resonances can be found in the normal leads, depending on the 
length $L_S$ of the superconductor region.
When a high gate potential is selected, 
the gate also acts as an effective electrical {\em cutoff}, interrupting any transmission between the two terminals. In this regime,  
the distance between the gate and the superconductor,
denoted by $\ell$,
plays a significant role in the chiral Majorana phase of the superconductor. 
An interference of chiral Majoranas manifests as an oscillation
in the
conductance 
of the right terminal
between 0 and 2$e^2/h$ as a function 
of $\ell$. 
This interference occurs due to the chiral Majorana injection, propagation, and reflection in the normal intermediate region.
In an experiment, one could vary $\ell$ in a single device
by activating one of a sequence of gates at a particular distance.

\section{Model}
\label{sec1}

\subsection{Hamiltonian}
\label{ham}

We consider an MTI-based QAH thin slab 
containing a region of proximity-induced superconductivity.
Another finite sector is under the influence of electrostatic top and bottom  gates (see Fig.\ \ref{F0}). The electrostatic gates are assumed to have reversed 
potentials such that a vertical electric field is induced in the slab while its mean potential remains zero. Thus, the  average
chemical potential in the gated sector is not shifted with respect to the other parts of the slab. 
It is also important to note that the gated sector breaks mirror symmetry (left-right) with respect to the
proximitized sector. We will discuss below how this manifests 
itself on the electrical conductances.

The system Hamiltonian reads \cite{Qi2010,Wang2015,Osca2018}
\begin{eqnarray}
\label{eq1}
{\cal H} &=& 
\left[\, m_0 + m_1 \left(p_x^2 +p_y^2\right)\, \right] \lambda_x\,\tau_z  
\nonumber\\
&-& \frac{\alpha}{\hbar}\, \left(\,p_x\sigma_y-p_y\sigma_x\right)\, 
\lambda_z \,\tau_z
+ \Delta_Z\, \sigma_z
-\mu\,\tau_z
\nonumber\\
&+& \left( \Delta_p + 
\Delta_m\,\lambda_z\right) \tau_x\,
+\Delta_g\,\lambda_z\,\tau_z\; ,
\end{eqnarray}
where $\sigma_{xyz}$, $\lambda_{xyz}$ and 
$\tau_{xyz}$ represent Pauli matrices for spin, layer and electron-hole
spaces, respectively.
The parameters $m_0$ and $m_1$ model the coupling between layers. 
$\alpha$ is the strength of the Rashba-like spin-orbit interaction,
and
$\Delta_Z$ is the Zeeman-like magnetization strength.
The superconductivity parameters $\Delta_{p,m}$ are linear combinations of the pairing strengths  
in each layer. Specifically, we have that $\Delta_{p,m}\equiv (\Delta_1\pm\Delta_2)/2$, with $\Delta_1$ and $\Delta_2$ the upper and lower layer pairings. Typically, these two parameters differ
because
the  s-wave superconductor lies closer to one of the layers, with the opposite layer receiving fewer Cooper pairs. Furthermore, it is well known that for a chiral Majorana phase these two values must differ \cite{Wang2015}.

The last term in Eq.\ (\ref{eq1}) models the electrostatic-gating effect
by means of the energy parameter $\Delta_g$, which has 
opposite signs on the two layers due to the $\lambda_z$ operator.
The physical origins of the various parameters in the bilayer model have been widely discussed in literature  \cite{Qi2010,Wang2013,Wang2014,Wang2015,Lian2016,Osca2018,Wang2015b,Wang2016,Chong2023}.
As mentioned, the above bilayer model model has three types of two-valued pseudospins 
corresponding to a Nambu spinorial basis
$\left[
(
\Psi^t_{k\uparrow},
\Psi^t_{k\downarrow},
\Psi^{t\dagger}_{-k\downarrow},
-\Psi^{t\dagger}_{-k\uparrow}
),
(
\Psi^b_{k\uparrow},
\Psi^b_{k\downarrow},
\Psi^{b\dagger}_{-k\downarrow},
-\Psi^{b\dagger}_{-k\uparrow}
)\right]^T
$.
We use the above Hamiltonian to describe junctions 
sketched in Fig.\ \ref{F0} by assuming that parameters $\{m_0,m_1,\alpha,\Delta_Z,\mu \}$ are constant throughout all sectors of the slab while the superconductivity $\Delta_{p,m}$ and 
gating $\Delta_g$ are nonvanishing only in the S and Ng regions, respectively.

Depending on the values of the magnetization,
the S region of the slab may be in one of three different phases. 
These are the trivial phase ${\cal N}=0$, a Majorana topological phase ${\cal N}=1$, and a phase with two Majoranas ${\cal N}=2$, where ${\cal N}$ is a Majorana topological number.
This
corresponds to the number of chiral Majorana modes propagating along the edges of the device. 
We refer to the literature for detailed studies of the phase diagram with uniform parameters; particularly Refs.\ \cite{Wang2015,Wang2015b,Legend2024}. 
Chiral Majoranas appear in the  energy band diagrams 
$\varepsilon(k)$ as 
non degenerate
Dirac-like crosses 
inside the superconducting energy gap. The crossing point occurs at $k=0$ and $E=0$.

We use a numerical method to calculate the eigenfunctions 
within each device region,
and a matching algorithm to determine the outgoing amplitudes
for a given input from the asymptotic leads.
Our unit system uses $1\,{\rm meV}$ as the energy unit and  
$1\,\mu{\rm m}$ as the length unit.
Unless stated otherwise, in this paper we
assume the following values for the Hamiltonian parameters:
$\alpha = 0.2\, {\rm meV}\mu{\rm m}$, $m_0=17\, {\rm meV}$, 
$\hbar m_1=10^{-3}\,{\rm meV} \mu{\rm m}^2$,  
$\Delta_{1}= 1\, {\rm meV}$ and $\Delta_2=0$. 
The ferromagnetic magnetization parameter, $\Delta_Z$, 
will be tuned
to demonstrate its impact on the conductance of the various superconductor phases. This parameter 
physically
models the intrinsic magnetization of the material.
The above Hamiltonian parameters are taken as reasonable estimates for QAH insulators based on Cr doped and V doped Bi$_2$Se$_3$ magnetic 
thin films as discussd in Refs.\ \cite{Legend2024,Zsurka2024}.
Nevertheless, 
the results discussed below are not tied to this particular set of parameters.

\subsection{The scattering formalism}
\label{scatt}

The probabilities of the device's different outgoing modes 
are calculated using a complex band structure analysis (see, e.g., Refs.\ \cite{Osca2019,DiMi2023}). 
In each region of the junction, the solution is a linear superposition of left-going, right-going and evanescent modes, given by
\begin{equation}
\label{eq2}
\Psi (x,y) = \sum_{k_i \sigma\tau\lambda} 
C_{k_i}\,
\Phi^{(k_i)}_{\sigma\tau\lambda}(y)\;
e^{i k_i x}\; ,
\end{equation}
where the $k_i$ correspond to the set of complex wave numbers compatible with a total energy $E$.
The $C_{k_i}$'s are the set of complex amplitudes in each region,
and the  
$\Phi^{(k_i)}_{\sigma\tau\lambda} (y)$'s are the eight component 
spinors ($\sigma,\tau,\lambda=1,2$) for each complex wavenumber $k_i$.
Right-going and left-going propagating modes are characterized by positive and negative real wavenumbers, respectively.
Complex (i.e., non real) wavenumbers represent evanescent modes. 

In practical terms, 
the set of wave numbers and eigenspinors  $\{k_i, \Phi^{(k_i)}\}$ is obtained by
diagonalizing the $k$-eigenvalue problem of a homogeneous infinite slab with the same parameters as the region,
for each slab sector.
The values of $\{ C_{k_i}\}$ are then given by
a linear system 
representing a
generalized matching algorithm
at the junction interfaces. Further details of the complex band structure method can be found in  Appendix \ref{appA}, as well as in previous Refs.\ 
\onlinecite{Osca2019,DiMi2023}.

Using the outgoing probability amplitudes, we can 
calculate the four elements of the conductance matrix,
\begin{equation}
\left(\begin{matrix}
I_L \\
I_R
\end{matrix} \right)
=\left(\begin{matrix}
 g_{LL} & g_{LR} \\
 g_{RL} & g_{RR} 
\end{matrix} \right)
\left(\begin{matrix}
V_L \\
V_R
\end{matrix} \right)\; ,
\end{equation}
where $I_{L/R}$ and $V_{L/R}$  are the current and potential at the left and right leads, respectively, and  $g_{ij}=\partial I_i / \partial V_j$ is the electrical conductance matrix for the different leads, $i/j=L/R$. 
The electric current $I_i$ at the QAH terminals can be computed as \cite{Lamb1993}
\begin{equation}
\label{eq4}
I_i = \int_0^\infty dE\, \sum_p s_p\, \left[J_i^p (E)-K_i^p (E)\right]\; ,
\end{equation}
where $p \in {e,h}$, with $e$ standing for electron and $h$ standing for hole, and $s_{e/h}=\pm 1$.

In Eq.\ (\ref{eq4}) the incident current $J_i^p (E)$  is calculated as
\begin{equation}
J_i^p(E) =  \frac{e}{h}\,   N_i^p(E)\, f_i^p (E,V_i,T_i)\; ,
\end{equation}
where $N_i^p (E)$ is the number of propagating modes of kind $p$ in lead $i$, 
and 
$f_i^p \equiv 1/[ 1+\exp(E-s_p e V_i)/kT_i ]$
is the corresponding Fermi distribution.
The lead temperatures $T_i$  
will be assumed to be negligible.  
Conversely, the outgoing current $K_i^p(E)$ in Eq.\ (\ref{eq4}) 
is given by
\begin{equation}
K_i^p(E) =  \frac{e}{h}\, \sum_{j,q} P_{ij}^{pq}(E)\, f_j^q (E,V_j,T_j)\; ,
\end{equation}
where $P_{ij}^{pq}$ represents the transition probability of a quasiparticle of kind $q$ at lead $j$ to a quasiparticle of kind $p$ at lead $i$. 

Note that equal indexes $i=j$ imply reflection on 
the same terminal while different ones $i \neq j$ imply transmission
between terminals. 
Also note that electron and hole quasiparticle numbers are not  conserved 
independently
at superconductor interfaces. 
Reflection and transmission may involve
the same kind of quasiparticle or a quasiparticle swap. 
For example, $P_{ii}^{he}$ is the probability of Andreev electron-hole reflection in which the number of electrons changes by two due to the formation of a Cooper pair. 
The resulting expressions for the conductances at vanishing temperature are
\begin{eqnarray}
\label{eq7}
g_{LL}&=&N_L^e - P_{LL}^{ee} + P_{LL}^{he}\; , \\
\label{eq8}
g_{RR}&=&N_R^e - P_{RR}^{ee} + P_{RR}^{he}\; , \\
\label{eq9}
g_{RL}&=& P_{RL}^{he} - P_{RL}^{ee}\; ,\\
\label{eq10}
g_{LR}&=& P_{LR}^{he} - P_{LR}^{ee}\; .
\end{eqnarray}
In Eqs.\ (\ref{eq7})-(\ref{eq10}) all quantities are evaluated at an energy
$E=e\,V_j$, with $j$ the corresponding terminal index of the 
conductance matrix $g_{ij}$.

\section{Results}
\label{sec3}

\begin{figure}[t]
  \centering
\includegraphics[width=0.5\textwidth,trim=1cm 7.5cm 2.5cm 2cm,clip]{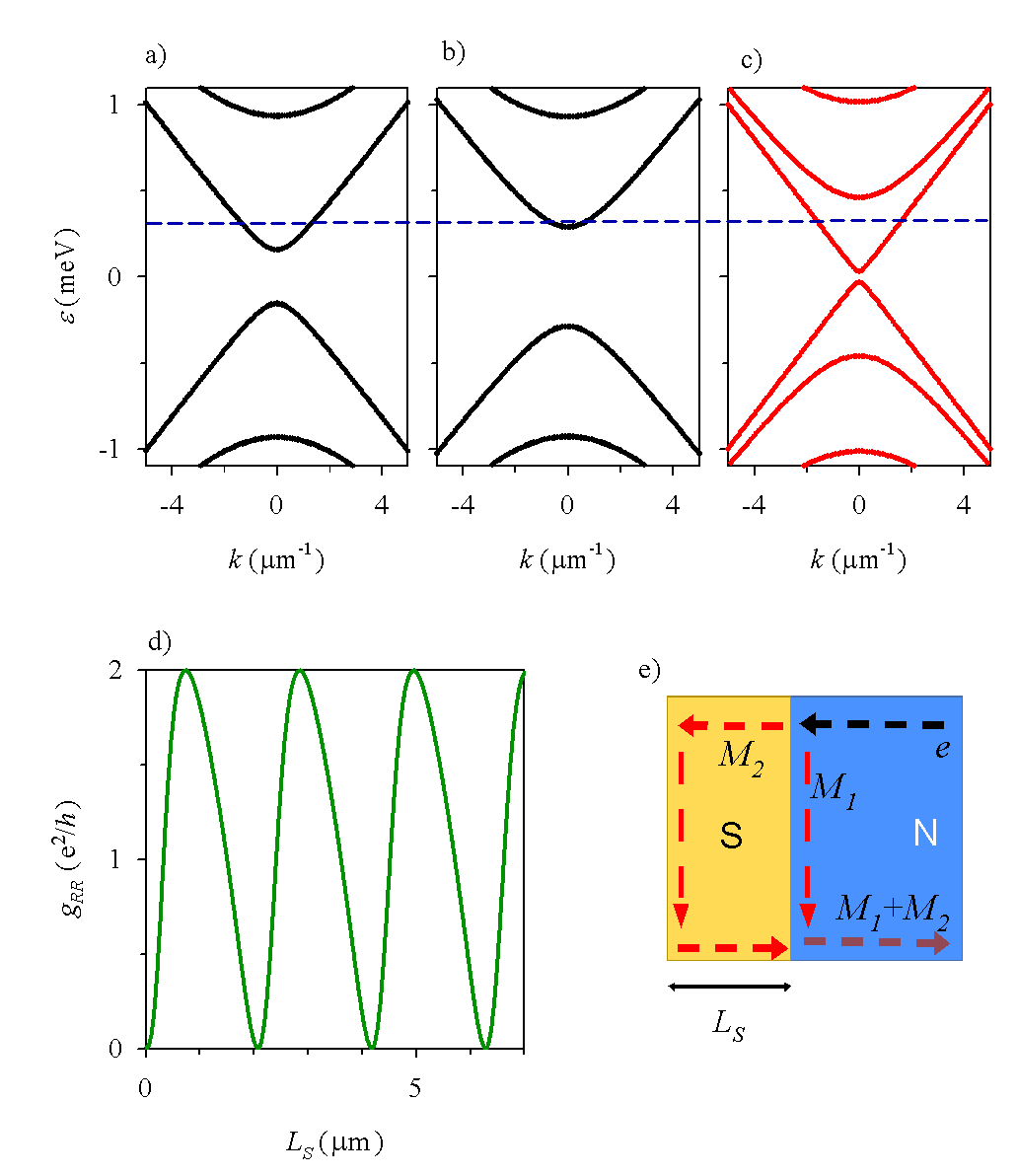}
\caption{ a,b,c) Band structure  in
regions N (a), Ng (b) and S (c) of the slab sketched in Fig.\ \ref{F0}. We used 
$L_y=1\,\mu{\rm m}$, $\Delta_Z=17.3 \,{\rm meV}$
and
$\Delta_g=3\, {\rm meV}$ such that 
for $E=0.3\, {\rm meV}$ (dashed horizontal line) region Ng is becoming gapped.
d) Local conductance in the right lead as a function of the superconductor length $L_S$ when an intense electrostatic gating $\Delta_g$ is applied, effectively disconnecting L and R 
terminals. 
e) A schematic of the resulting interference effect when a right-incident QAH  electron mode is effectively reflected into two interfering Majoranas $M_1$ and $M_2$ in lead R. }
  \label{F2}
\end{figure}

In this Section, we present the results of the paper on electrical 
properties of the system and how they are affected by the presence of the electrostatic gate. In Fig.\ \ref{F2} we show the band structure of the normal leads (Fig.\ \ref{F2}a), the gated region with an electric field (Fig.\ \ref{F2}b) and the region in contact with a superconductor (Fig.\ \ref{F2}c). Overall, the electric field has two effects: it widens the energy gap and flattens the bands. 

The widening of the gap induced by the gating in the Ng region
is analogous to the effect of reducing the width $L_y$ of the wire.  
Conversely, the S region, in contact with a superconductor,
works as  a topological superconductor.  
Depending on the pairings $\Delta_{1,2}$ and 
magnetization $\Delta_Z$, it can be in different phases labeled as $\mathcal{N}=0$, $\mathcal{N}=1$ and 
$\mathcal{N}=2$ \cite{Wang2015b}. 
Figure\ \ref{F2}c shows the typical band structure for 
the $\mathcal{N}=1$ phase, in which a pair of Majorana propagating modes are present. One has $k>0$ and the other has $k<0$. Both have a linear dispersion relation, with each propagating in opposite edges of the device. In the $\mathcal{N}=2$  phase two Majoranas propagate along each edge instead of one. Therefore two linear dispersion bands appear for $k>0$ and $E>0$ (see Appendix \ref{appB} and Fig.\ \ref{F10}).
In an infinitely wide system this new pair of bands is degenerate but finite-size effects break the perfect degeneracy. The trivial phase, the $\mathcal{N}=0$ phase, has no propagating bands inside the gap 
in S.

The propagating modes in the normal regions N and Ng are doubly degenerate
in Nambu space, for electrons and holes. 
Here, the charge, labelled by the quantum number $\tau_z=\pm 1$ is a proper quantum label due to the 
absence of any pairing $\Delta_{pm}=0$. By contrast, the chiral Majoranas of the proximitized S region are chargeless and are equal superpositions of electrons and
holes. In the limit of vanishing chemical potential, $\mu\approx 0$, 
the neutral character of the Majoranas is maintained over a broad 
range of $E$ and $k$ values (Fig.\ \ref{F2}c). However, when $\mu$ starts 
to deviate from zero, this range shrinks to the vicinity of vanishing
$E$ and $k$ values. For this reason chiral Majorana physics is better observed in the 
$\mu\approx 0$ limit due to the finite size effect.  
In Sec.\ \ref{nonloc} we discuss the initial 
departures from this regime that were observed in the nonlocal conductances. 

\subsection{Large $\Delta_g$ limit}

It can be seen in Fig.\ \ref{F2}b that 
with energy $E$
there is a critical electrostatic gating $\Delta_g$ that, when exceeded, renders conduction
in region Ng of the slab suppressed due to the absence of propagating modes.
We refer to this as the cutoff regime of the gated slab. 
Under these conditions, the conductance $g_{LL}$ is naturally zero at low energies.
This situation is equivalent to cutting the wire electrically. 
However, $g_{RR}$ still oscillates between zero and two 
quantum units, 
depending on the length of the 
${\cal N}=1$ superconductor region
(Fig.\ \ref{F2}d). 
A Fermion QAH state traveling from right to left will 
encounter a topological superconductor interface. 
As shown in Fig.\ \ref{F2}e, 
half of the mode will be reflected as a Majorana $M_1$ 
on this interface,
while the other half will be transmitted as a Majorana $M_2$. 
When this last Majorana quasiparticle turns around at the virtual left edge due to the discontinuity created by the electric field, it will be reflected back to its original right lead. Both Majoranas will now run attached to the lower 
edge of the right lead and interfere which each other in a constructive or destructive manner, depending on their relative phase difference, which varies with the superconductor length $L_S$. 
A similar interference pattern can be obtained with a fixed superconductor length $L_S$ but
varying the energy $E$, which is equivalent to varying the applied right bias $V_R$.
We stress that this remarkable interference mechanism only emerges on the right lead when the gating $\Delta_g$ is large.
The variation with smaller values of the electrostatic gating is discussed 
next.

\subsection{Local conductances}

\begin{figure}[t]
  \centering
\includegraphics[width=0.4\textwidth,trim=1cm 0.5cm 2cm 1.5cm,clip]{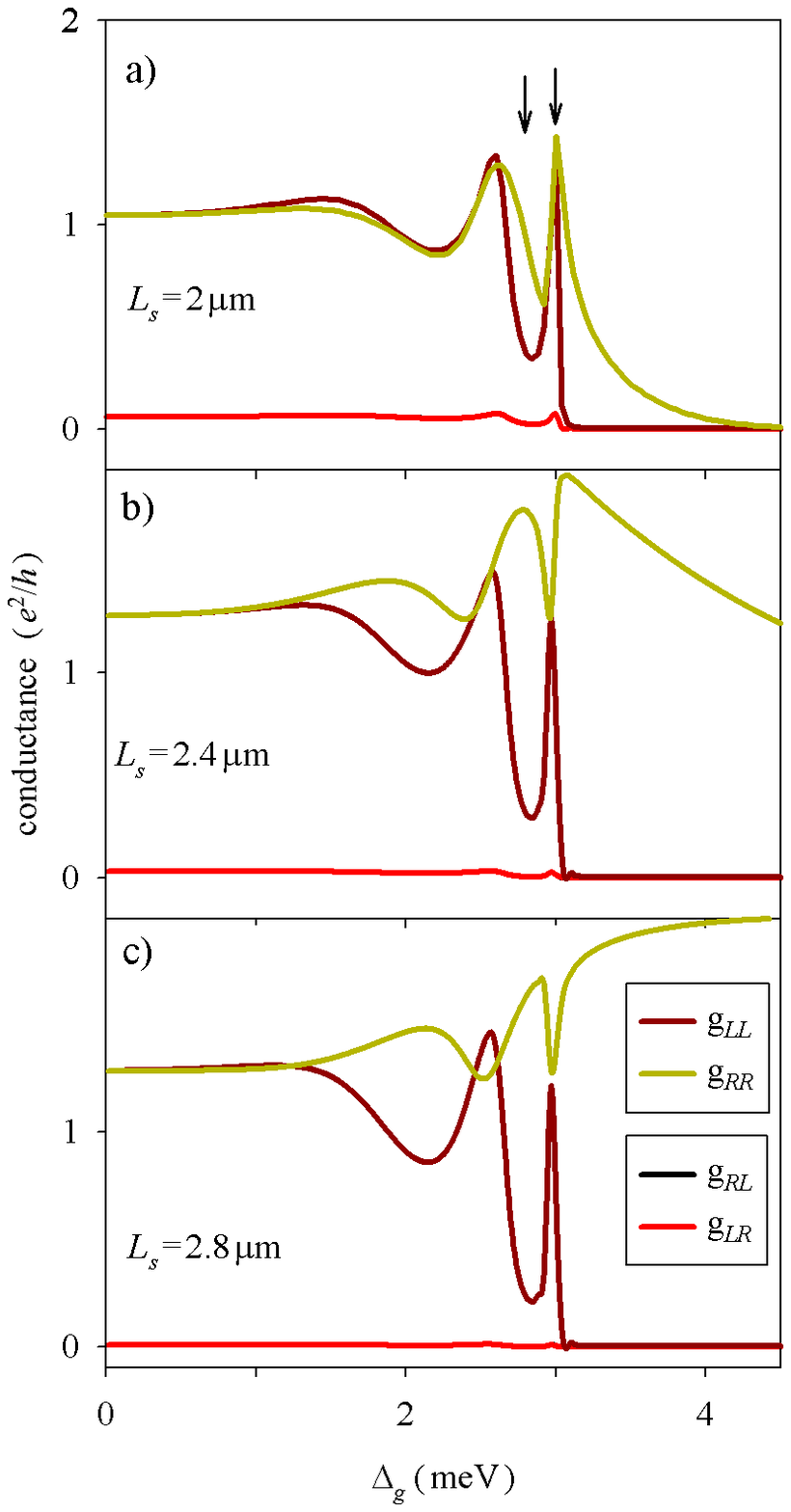}
\caption{Local conductances as a function of the strength of electrostatic gating $\Delta_g$. The results in each panel correspond to a different length $L_S$ of the superconducting region, as indicated in the figure. We assumed $L_y=1\; \mu{\rm m}$,
$L_g=7\; \mu{\rm m}$, $\ell\approx 0$, $E=0.3\; {\rm meV}$, $\Delta_Z=17.3\;{\rm meV}$,
and the other parameters 
are given in Sec.\ \ref{ham}. 
}
  \label{F3}
\end{figure}

The local conductances $g_{LL}$ and $g_{RR}$ are displayed
in Fig.\ \ref{F3}, showing how they change with
electric field strength for three different lengths of the superconductor.
Energy and magnetization correspond to the 
${\cal N}=1$ phase of the superconductor. 
Each superconductor length represents a different scenario, with 
conductance peaks and valleys exactly correlated, anti-correlated or somewhere in between.
These three scenarios can be distinguished by the value of $g_{RR}$ 
for large  $\Delta_g$. Conversely, the $g_{LL}$ dependence
is not greatly affected by the 
superconductor length $L_S$. 

When $g_{RR}\to 2e^2/h$ for large $\Delta_g$, peaks and valleys in $g_{LL}$ and $g_{RR}$ are anti-correlated across the entire range of gating intensities (Fig.\ \ref{F3}c). 
However, when $g_{RR}\to 0$ for high $\Delta_g$, $g_{LL}$ 
and $g_{RR}$ are nicely correlated (Fig.\ \ref{F3}a). Intermediate scenarios are realized for intermediate values of $g_{RR}$ after the cutoff
at large values of $\Delta_g$. 
The overall behavior can be explained by realizing that as the Ng region becomes increasingly opaque $g_{LL}$ eventually vanishes after 
several peaks and valleys. However, $g_{RR}$ may eventually decrease or increase due to the Majorana interference mechanism discussed above.
Therefore, Ng electrostatic gating clearly manifests the Majorana modes as a function 
of $\Delta_g$.
Note that the manner in which the Ng-S interface becomes opaque is not straightforward, a topic we will expand on next.

\begin{figure}[t]
  \centering
\includegraphics[width=0.45\textwidth,trim=0cm 10cm 0.5cm 2.cm,clip]{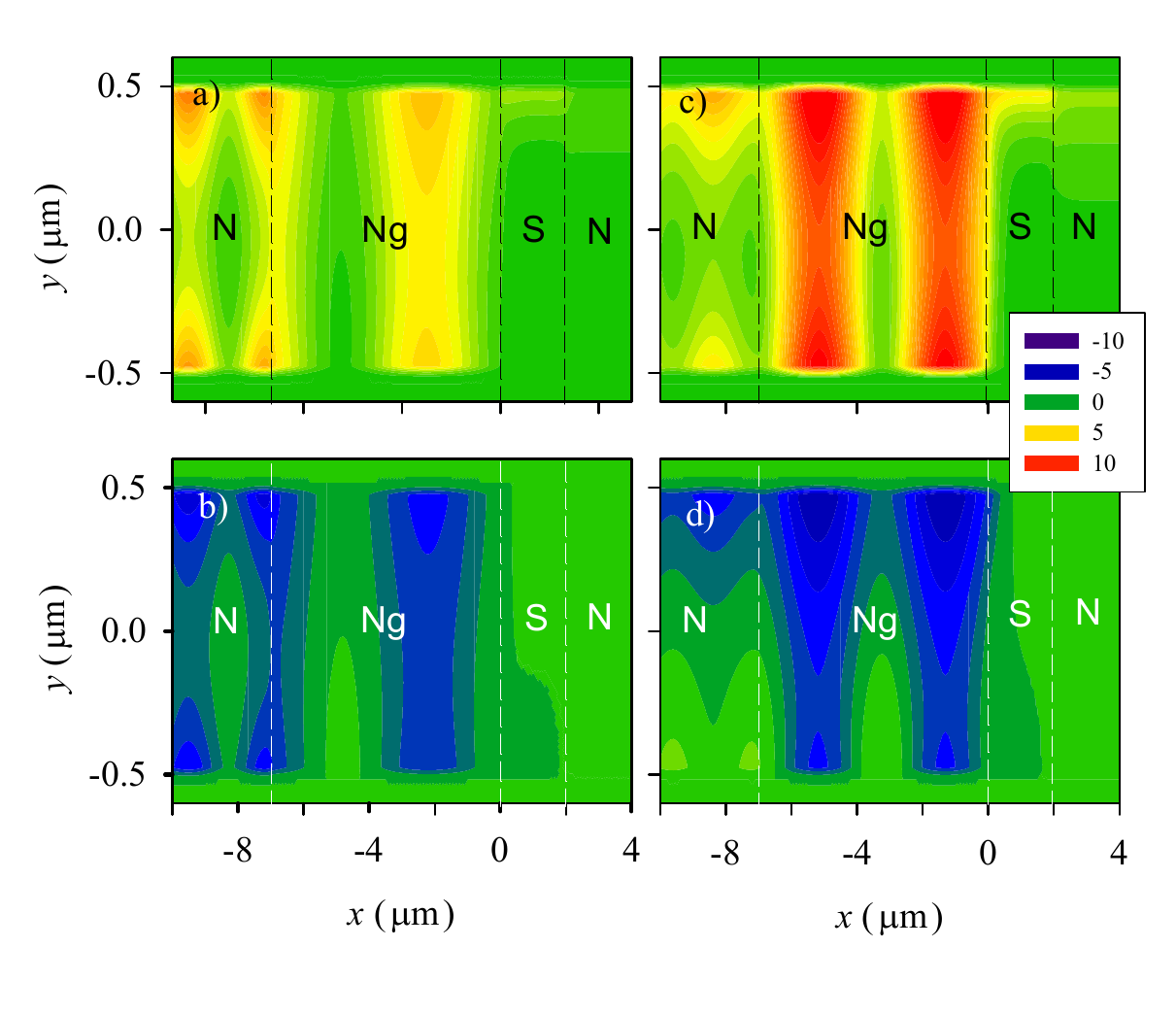}
  \caption{Probability density distributions (a, c) and charge density 
distributions (b, d) in the various junction regions
corresponding to a left-incidence electron mode
when 
$L_g=7\; \mu{\rm m}$ and $L_S=2\; \mu{\rm m}$, as in Fig.\ \ref{F3}a. 
The strength of the electric field is $\Delta_g=2.85\; {\rm meV}$
in panels a) and b), and  $3\; {\rm meV}$
in panels c) and d). These two values of $\Delta_g$ correspond to 
the conductance dip and peak
indicated with
arrows in 
Fig.\ \ref{F3}a.}
  \label{F5}
\end{figure}

When $\Delta_g\to 0$ in Fig.\ \ref{F3} both $g_{LL}$ and $g_{RR}$ are close to $e^2/\hbar$. However, the actual values may differ slightly due to the partial reflection of Majoranas caused by the finite size of 
the superconducting region. 
In the gated
Ng region, the effects of finite size are greatly enhanced by increasing the value of $\Delta_g$. 
Figure \ref{F5} shows the spatial distributions of the quasiparticle probability and charge for two different gating strengths, $\Delta_g$.
In the presence of an external electric field, it is possible to create low-wavenumber bulk states in the gated
Ng region. These states result from the hybridization of a left-incident Fermionic QAH mode and a partially reflected Majorana mode at the superconductor interface. It is these low wavenumber states that are responsible for the wide oscillations in conductance 
seen as peaks and deep valleys in $g_{LL}$ and $g_{RR}$
with increasing $\Delta_g$. 
The characteristic long wavelengths of these states are due to the flattening of their transport bands. The presence of these states
is not exclusive of superconducting topological junctions. 
Similar quasibound  states can also be obtained in gated sections of normal QAH slabs. However, the main difference is that, with 
superconductivity, the conductance peaks can rise well above the quantum of conductance.

When an  integer number of half wavelengths fit within the length $L_g$ of the gated Ng region,
there is a peak in $g_{LL}$ (see Fig.\ \ref{F5}c,d). Conversely, 
if this number is a semi-integer, a conductance valley is observed (Fig.\ \ref{F5}a,b).  From the charge distributions of these states,
we can see that, in Fig.\ \ref{F5}b,
an electron from the upper left is 
reflected also as an electron in the lower left, thus
the minimum in $g_{LL}$. 
On the other hand, Fig.\ \ref{F5}d shows a vanishing 
electron reflection, or even a slight Andreev positive charge 
reflection, causing a maximum in $g_{LL}$ conductance.

\begin{figure}[t]
  \centering
\includegraphics[width=0.45\textwidth,trim=1cm 15cm 1cm 2cm,clip]{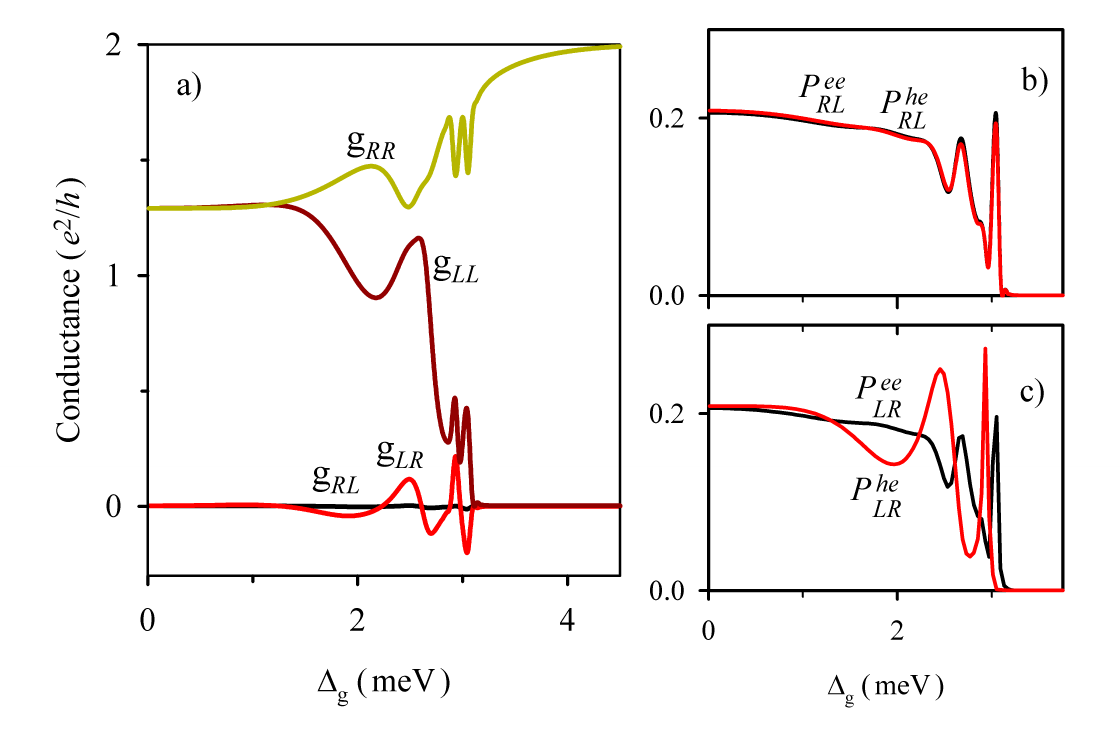}  \caption{a) Conductances  
similar to Fig.\ \ref{F3}c but with a
chemical potential $\mu=5{\times}10^{-3}\,{\rm meV}$.
The corresponding $L{\to}R$ and $R{\to}L$ transmission probabilities are shown in panels b and c, respectively.
The 
$L{\to}R$ transmissions for electrons and holes are identical (b).
Differences are clearly visible for
$R{\to}L$ (c). This manifests as a directional 
asymmetry in the nonlocal conductances with a non-vanishing $g_{LR}$,
as
shown in panel a.
}
\label{F4}
\end{figure}

\subsection{Non-local conductances}
\label{nonloc}

Non-local conductances offer an alternative view.
In Fig.\ \ref{F3}, both $g_{LR}$ and $g_{RL}$ are effectively 
vanishing over the entire
range of gate intensities. This can be understood by noting that
in the limit of $\mu\approx 0$ quasiparticle transmissions between 
terminals are identical and independent of particle type; i.e.,
$P_{RL}^{ee}=P_{RL}^{he}=P_{LR}^{ee}=P_{LR}^{he}$.
Deviations from this behavior
when slightly increasing $\mu$
are first seen in the nonlocal conductances, as shown in Fig.\ \ref{F4}. 
Even a rather small $\mu$ is enough to yield sizeable nonlocal conductance oscillations
in $g_{LR}$ (Fig.\ \ref{F4}a), 
while $g_{LL}$, $g_{RR}$ and $g_{RL}$ remain in the $\mu\approx 0$ scenario.
Nonvanishing $g_{LR}$ and vanishing $g_{RL}$ correspond to the 
emergence of a Majorana 
diode effect, which is tunable via electrostatic gating
\cite{Nade2023}.

The physics in Fig.\ \ref{F4} can be understood from 
the asymmetry of the device. A right-going QAH edge state
from the left terminal first traverses the electric field, followed by the topological superconductor.
In contrast, a left-going edge state traverses the regions in reversed order.
The probabilities of a chiral Majorana S$\,\to\,$N injecting  an electron or a hole are identical  (Fig.\ \ref{F4}b). 
However, these probabilities differ 
for a chiral Majorana S$\,\to\,$Ng injection (Fig.\ \ref{F4}c).
This asymmetry is highly sensitive to the chemical potential $\mu$.
It vanishes in the strict $\mu=0$ limit, but yields 
sizeable deviations in the non-local conductances at rather
small values of $\mu$ that do not gretly affect the local coductances.

\begin{figure}[t]
\centering
\includegraphics[width=0.45\textwidth,trim=1cm 16cm 1cm 2cm,clip]{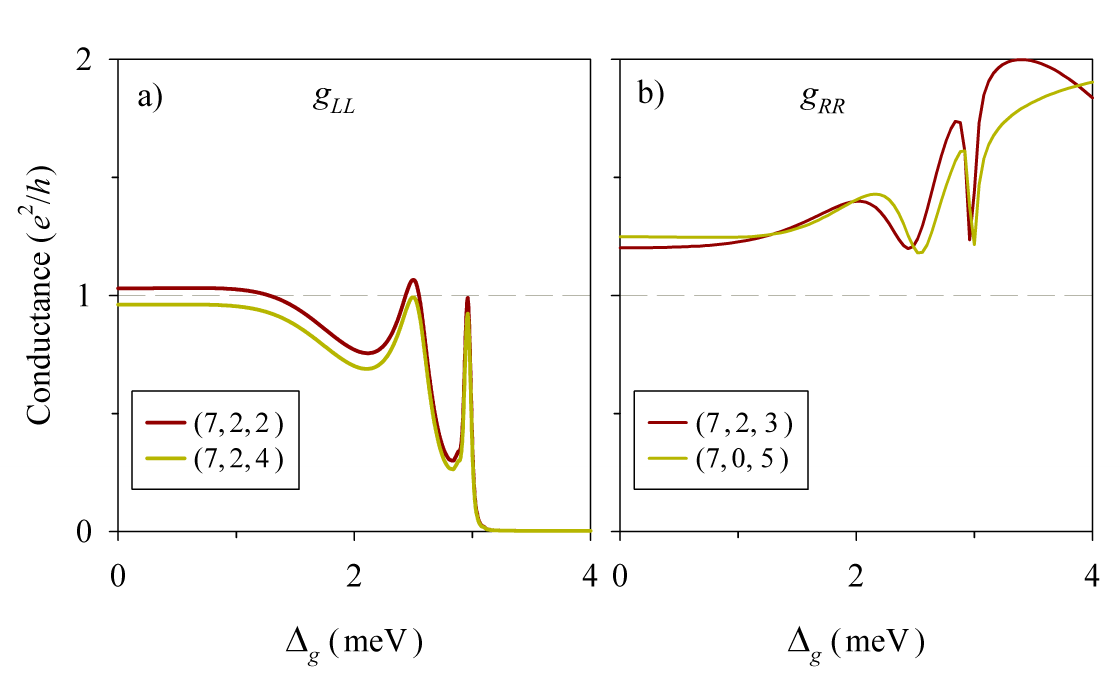}  
\caption{Local conductances with nonvanishing Ng-S distance $\ell$. 
The figure keys provide the distances $(L_g,\ell,L_S)$ in microns.
The rest of parameters are the same as in Fig.\ \ref{F3}.
}
\label{F6}
\end{figure}

\subsection{Role of Ng-S spacing $\ell$}

In this section,
we consider  the role of the distance  between the normal gated (Ng)  and proximitized superconducting (S) sectors of the MTI slab, 
labeled as $\ell$ in Fig.\ \ref{F0}.  
In all previous results, we assumed that this distance was 
approximately zero.
Here, we focus on how the presence of an  intermediate normal sector  (see Fig.\ \ref{F0}) modifies the local conductances $g_{LL}$ and $g_{RR}$. 
We find that, for $\mu\approx 0$, the non-local conductances vanish again in the presence of this distance .

The role of the distance $\ell$ is different for the left $g_{LL}$ and right 
$g_{RR}$ conductances. In fact, when all three distances 
$(L_g, \ell, L_S)$ are considered, the left conductance $g_{LL}$ is not very sensitive to $L_S$. As shown in Fig.\ \ref{F6}a the overall dependence on the electrostatic gating, especially the resonant peaks before the 
effective cutoff for $\Delta_g\gtrsim 1$ meV, is qualitatively similar
for fixed $L_g$ and $\ell$, but with different $L_s$ values. This is 
reasonable in view of the quasi-confined state mechanism discussed above, which  essentially depends on the left incidence
conditions, which are the relevant  ones for $g_{LL}$.

For the right terminal conductance ($g_{RR}$)
we find that systems with a given $L_g$ and a given value of $\ell+L_S$
show very similar results, even when $L_S$ and $\ell$ differ (Fig.\ \ref{F6}b). 
This effective dependence
on $\ell+L_S$ can be understood in terms of the 
propagation of a chiral Majorana. Indeed, a left-propagating chiral Majorana injected from S into the intermediate normal region
of length $\ell$ preserves its Majorana character also in 
the normal region. Therefore, in the presence of a chiral Majorana in S, $g_{RR}$
does not distinguish between
the systems with $(\ell,L_S)$ and $(\ell'=0,L_S'=L_S+\ell)$.
As shown in Fig.\ \ref{F6}b, this qualitative similarity holds true 
for all gating intensities, including the resonance
peaks and the large conductance in the wire cutoff regime.  

The distance $\ell$ provides a way to mimic the chiral Majorana 
$g_{RR}$ conductance of a longer proximitized sector $L_S$ with
a shorter one $L'_S$, such that
$L_S=L'_S+\ell$. This can be important since the coherence length
of Cooper pairs 
in the superconductor that gives rise to the proximity effect in the S region
is usually much shorter than that of the QAH edge modes. Therefore, 
an intermediate slab sector of distance $\ell$ can be a 
practical way 
to avoid the need for very large values of $L_S$ while still
retaining the chiral Majorana quantum coherence.

\subsection{A chiral Majorana protocol}

\begin{figure}[t]
\centering
\includegraphics[width=0.45\textwidth,trim=0.5cm 16cm 0.5cm 2cm,clip]{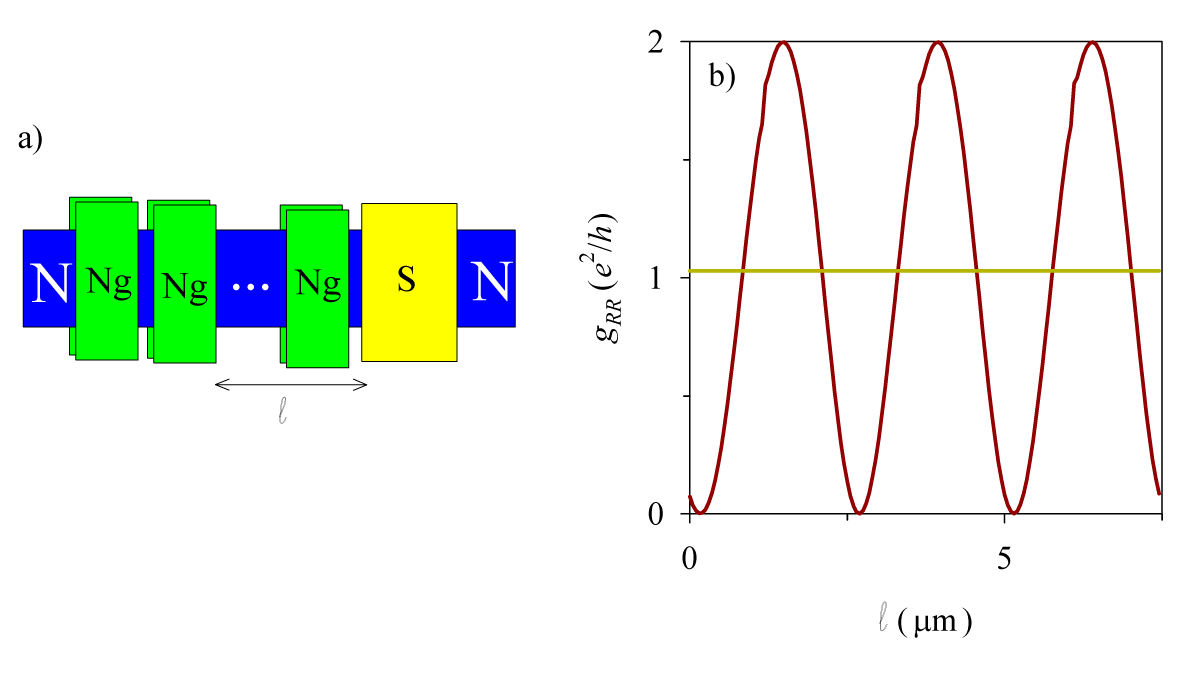}  
\caption{
a) Sketch of an MTI slab (blue) proximitized on top with a superconductor (yellow) and with a sequence of top-bottom electrostatic
gates (green) to the left of the superconductor.
b) Dependence of $g_{RR}$ with the distance $\ell$ 
between the superconducting region and
the electrode 
that electrically cuts the wire. 
The horizontal line shows the 
reference value $g_{RR}=e^2/h$ when all the electrodes are set to zero. Parameters: $E=0.3$ meV, $\Delta_Z=17.3$ meV, $\Delta_g= 7$ meV, $L_g=7$ $\mu$m, $L_S=2$ $\mu$m. 
}
\label{F7}
\end{figure}

The above results suggest a protocol for identifying the presence of a
chiral Majorana mode in a proximitized MTI slab, based on a sequence of 
multiple side electrical gates (Fig. \ref{F7}).
By applying a large $\Delta_g$ to a selected gate, while keeping the others at zero gating, one can perform an effective wire cutoff at 
different distances from the proximitized superconducting region.
Importantly, this can be achieved purely by electrical means with a single device, elliminating the need to fabricate  multiple devices of different lengths of proximitized superconductor from scratch.

Specifically, we propose the following two-step protocol 
to identify the chiral Majorana scenario: 
\begin{itemize}
\item[a)] With $\Delta_g=0$ in all gates, $g_{LL}\approx g_{RR}\approx e^2/h$; 
\item[b)] With a large $\Delta_g$ in a selected
single gate, $g_{LL}\approx 0$ while $g_{RR}$ takes values between 0 and $2 e^2/h$ depending on the distance $\ell$ of that particular gate to the S sector. 
\end{itemize}
Condition a) is equivalent to the original chiral Majorana signature discussed 
in Ref.\ \cite{Wang2015}. While a) is indeed a necessary condition for chiral Majoranas, it was later shown to be  insufficient, as it can be mimicked in a non Majorana scenario
where an incident QAH edge mode equilibrates in the superconductor 
emitting electron and holes with equal probabilities
\cite{Ji2018,Huang2018,Kay2020,Uday2025}. 
This mechanism would be ruled out by requirement b). In fact, this second requirement proves the quantum mechanical interference between the two chiral Majoranas $M_1$ and $M_2$ discussed in Fig.\ \ref{F2} and therefore it
provides a much more solid complementary evidence.
Interferometric conditions in a different setup, with a circular Corbino geometry
were discussed in Ref.\ \cite{Lian2018}.

Note that the above interferometric behavior 
differs from that
suggested in Ref.\ \cite{Osca2018}, where reducing the distance $L_y$ led to interference between the two edges. 
The present chiral Majorana conditions a) and b) still apply even for large $L_y$'s such that the 
two edges do not overlap significantly. The only requirement is the 
possibility of creating an
electric cutoff by electrostatic gating at varying distances of the proximitized region.
This can also be  viewed as probing the nonlocality of the extended 
chiral Majorana state, as it is affected by an operation (electrostatic gating) performed at a distance of the superconducting region.

\section{Conclusions}

\label{sec4}

Electrically gating a section of a QAH slab that also has a 
superconducting proximitized 
section offers interesting ways of probing the 
transport properties related to chiral Majorana physics.
A gap in the spectrum  opens locally in the gated part. This allows 
tuning of the local conductances
in the left and right terminals
by means of the gating strength
$\Delta_g$.  
In the Majorana phase of the proximitized slab,
resonances in the local conductances are observed.
These resonances are a 
function of the gating strength due to the resonant coupling with
localized states in the gated part of the slab. These local conductance 
resonances can exceed $e^2/h$ and they
are either correlated or anticorrelated  
depending on the length of the proximitized sector 
$L_S$, or more precisely on the total length $\ell+L_S$ where $\ell$ is the distance 
from the gated to the superconducting sectors.
The resonances can also be viewed as a function of energy (applied bias).
Nonlocal conductances usually vanish, but
when the chemical potential is increased from zero, an asymmetry
with nonvanishing right-to-left
nonlocal conductance emerges.
This mechanism has been suggested as
a gate-controlled Majorana diode effect on nonlocal conductances.

For strong gating, an electrical cutoff regime is found 
in which the left and right terminals
are effectively disconnected. In this regime, a chiral Majorana yields
large oscillations 
between 0 and $2e^2/h$ 
of the right local conductance
as a function of $\ell+L_S$, while all other conductances vanish.
Based on this, a protocol has been suggested to identify a chiral Majorana phase. This 
requires a proximitized QAH slab with multiple side electric gates. 
In the two-Majorana (Fermionic) phase
the scenario of a QAH mode propagating throughout the slab
is obtained: local conductances are identical, they do not exceed $e^2/h$ as a function  
of $\Delta_g$ and they tend to vanish in the electrical cutoff regime.

%\FloatBarrier

\begin{acknowledgments}

This project is financially supported  
by MCIN/AEI/10.13039/501100011033 under 
project PCI2022-132927 of the QuantERA grant MAGMA,
under project PID2023-151975NB-I00
and by the European Union NextGenerationEU/PRTR.
\end{acknowledgments} 

\bibliography{QAHn}

\begin{thebibliography}{46}%
\makeatletter
\providecommand \@ifxundefined [1]{%
 \@ifx{#1\undefined}
}%
\providecommand \@ifnum [1]{%
 \ifnum #1\expandafter \@firstoftwo
 \else \expandafter \@secondoftwo
 \fi
}%
\providecommand \@ifx [1]{%
 \ifx #1\expandafter \@firstoftwo
 \else \expandafter \@secondoftwo
 \fi
}%
\providecommand \natexlab [1]{#1}%
\providecommand \enquote  [1]{``#1''}%
\providecommand \bibnamefont  [1]{#1}%
\providecommand \bibfnamefont [1]{#1}%
\providecommand \citenamefont [1]{#1}%
\providecommand \href@noop [0]{\@secondoftwo}%
\providecommand \href [0]{\begingroup \@sanitize@url \@href}%
\providecommand \@href[1]{\@@startlink{#1}\@@href}%
\providecommand \@@href[1]{\endgroup#1\@@endlink}%
\providecommand \@sanitize@url [0]{\catcode `\\12\catcode `\$12\catcode
  `\&12\catcode `\#12\catcode `\^12\catcode `\_12\catcode `\%12\relax}%
\providecommand \@@startlink[1]{}%
\providecommand \@@endlink[0]{}%
\providecommand \url  [0]{\begingroup\@sanitize@url \@url }%
\providecommand \@url [1]{\endgroup\@href {#1}{\urlprefix }}%
\providecommand \urlprefix  [0]{URL }%
\providecommand \Eprint [0]{\href }%
\providecommand \doibase [0]{https://doi.org/}%
\providecommand \selectlanguage [0]{\@gobble}%
\providecommand \bibinfo  [0]{\@secondoftwo}%
\providecommand \bibfield  [0]{\@secondoftwo}%
\providecommand \translation [1]{[#1]}%
\providecommand \BibitemOpen [0]{}%
\providecommand \bibitemStop [0]{}%
\providecommand \bibitemNoStop [0]{.\EOS\space}%
\providecommand \EOS [0]{\spacefactor3000\relax}%
\providecommand \BibitemShut  [1]{\csname bibitem#1\endcsname}%
\let\auto@bib@innerbib\@empty
%</preamble>
\bibitem [{\citenamefont {Alicea}(2012)}]{Ali2012}%
  \BibitemOpen
  \bibfield  {author} {\bibinfo {author} {\bibfnamefont {J.}~\bibnamefont
  {Alicea}},\ }\bibfield  {title} {\bibinfo {title} {New directions in the
  pursuit of {M}ajorana fermions in solid state systems},\ }\href@noop {}
  {\bibfield  {journal} {\bibinfo  {journal} {Rep. Prog. Phys.}\ }\textbf
  {\bibinfo {volume} {75}},\ \bibinfo {pages} {076501} (\bibinfo {year}
  {2012})}\BibitemShut {NoStop}%
\bibitem [{\citenamefont {Beenakker}(2013)}]{Bee2013}%
  \BibitemOpen
  \bibfield  {author} {\bibinfo {author} {\bibfnamefont {C.~W.~J.}\
  \bibnamefont {Beenakker}},\ }\bibfield  {title} {\bibinfo {title} {Search for
  {{M}ajorana} fermions in superconductors},\ }\href
  {https://doi.org/10.1146/annurev-conmatphys-030212-184337} {\bibfield
  {journal} {\bibinfo  {journal} {Annu. Rev. Condens. Matter Phys.}\ }\textbf
  {\bibinfo {volume} {4}},\ \bibinfo {pages} {113} (\bibinfo {year}
  {2013})}\BibitemShut {NoStop}%
\bibitem [{\citenamefont {Elliott}\ and\ \citenamefont
  {Franz}(2015)}]{Elliot2015}%
  \BibitemOpen
  \bibfield  {author} {\bibinfo {author} {\bibfnamefont {S.~R.}\ \bibnamefont
  {Elliott}}\ and\ \bibinfo {author} {\bibfnamefont {M.}~\bibnamefont
  {Franz}},\ }\bibfield  {title} {\bibinfo {title} {Colloquium: {M}ajorana
  fermions in nuclear, particle, and solid-state physics},\ }\href
  {https://doi.org/10.1103/RevModPhys.87.137} {\bibfield  {journal} {\bibinfo
  {journal} {Rev. Mod. Phys.}\ }\textbf {\bibinfo {volume} {87}},\ \bibinfo
  {pages} {137} (\bibinfo {year} {2015})}\BibitemShut {NoStop}%
\bibitem [{\citenamefont {Sato}\ and\ \citenamefont
  {Fujimoto}(2016)}]{Sato2016}%
  \BibitemOpen
  \bibfield  {author} {\bibinfo {author} {\bibfnamefont {M.}~\bibnamefont
  {Sato}}\ and\ \bibinfo {author} {\bibfnamefont {S.}~\bibnamefont
  {Fujimoto}},\ }\bibfield  {title} {\bibinfo {title} {{M}ajorana fermions and
  topology in superconductors},\ }\href@noop {} {\bibfield  {journal} {\bibinfo
   {journal} {J. Phys. Soc. Jpn.}\ }\textbf {\bibinfo {volume} {85}},\ \bibinfo
  {pages} {072001} (\bibinfo {year} {2016})}\BibitemShut {NoStop}%
\bibitem [{\citenamefont {Schnyder}\ \emph {et~al.}(2008)\citenamefont
  {Schnyder}, \citenamefont {Ryu}, \citenamefont {Furusaki},\ and\
  \citenamefont {Ludwig}}]{Schny2008}%
  \BibitemOpen
  \bibfield  {author} {\bibinfo {author} {\bibfnamefont {A.~P.}\ \bibnamefont
  {Schnyder}}, \bibinfo {author} {\bibfnamefont {S.}~\bibnamefont {Ryu}},
  \bibinfo {author} {\bibfnamefont {A.}~\bibnamefont {Furusaki}},\ and\
  \bibinfo {author} {\bibfnamefont {A.~W.}\ \bibnamefont {Ludwig}},\ }\bibfield
   {title} {\bibinfo {title} {Classification of topological insulators and
  superconductors in three spatial dimensions},\ }\href@noop {} {\bibfield
  {journal} {\bibinfo  {journal} {Phys. Rev. B}\ }\textbf {\bibinfo {volume}
  {78}},\ \bibinfo {pages} {195125} (\bibinfo {year} {2008})}\BibitemShut
  {NoStop}%
\bibitem [{\citenamefont {Kitaev}(2009)}]{Kita2009}%
  \BibitemOpen
  \bibfield  {author} {\bibinfo {author} {\bibfnamefont {A.}~\bibnamefont
  {Kitaev}},\ }\bibfield  {title} {\bibinfo {title} {Periodic table for
  topological insulators and superconductors},\ }in\ \href@noop {} {\emph
  {\bibinfo {booktitle} {AIP conference proceedings}}},\ Vol.\ \bibinfo
  {volume} {1134}\ (\bibinfo {organization} {American Institute of Physics},\
  \bibinfo {year} {2009})\ pp.\ \bibinfo {pages} {22--30}\BibitemShut {NoStop}%
\bibitem [{\citenamefont {Qi}\ and\ \citenamefont {Zhang}(2011)}]{Qi2011}%
  \BibitemOpen
  \bibfield  {author} {\bibinfo {author} {\bibfnamefont {X.-L.}\ \bibnamefont
  {Qi}}\ and\ \bibinfo {author} {\bibfnamefont {S.-C.}\ \bibnamefont {Zhang}},\
  }\bibfield  {title} {\bibinfo {title} {Topological insulators and
  superconductors},\ }\href {https://doi.org/10.1103/RevModPhys.83.1057}
  {\bibfield  {journal} {\bibinfo  {journal} {Rev. Mod. Phys.}\ }\textbf
  {\bibinfo {volume} {83}},\ \bibinfo {pages} {1057} (\bibinfo {year}
  {2011})}\BibitemShut {NoStop}%
\bibitem [{\citenamefont {Sato}\ and\ \citenamefont {Ando}(2017)}]{Sato2017}%
  \BibitemOpen
  \bibfield  {author} {\bibinfo {author} {\bibfnamefont {M.}~\bibnamefont
  {Sato}}\ and\ \bibinfo {author} {\bibfnamefont {Y.}~\bibnamefont {Ando}},\
  }\bibfield  {title} {\bibinfo {title} {Topological superconductors: a
  review},\ }\href {https://doi.org/10.1088/1361-6633/aa6ac7} {\bibfield
  {journal} {\bibinfo  {journal} {Rep. Prog. Phys.}\ }\textbf {\bibinfo
  {volume} {80}},\ \bibinfo {pages} {076501} (\bibinfo {year}
  {2017})}\BibitemShut {NoStop}%
\bibitem [{\citenamefont {Prada}\ \emph {et~al.}(2020)\citenamefont {Prada},
  \citenamefont {San-Jose}, \citenamefont {de~Moor}, \citenamefont {Geresdi},
  \citenamefont {Lee}, \citenamefont {Klinovaja}, \citenamefont {Loss},
  \citenamefont {Nyg{\aa}rd}, \citenamefont {Aguado},\ and\ \citenamefont
  {Kouwenhoven}}]{Prada2020}%
  \BibitemOpen
  \bibfield  {author} {\bibinfo {author} {\bibfnamefont {E.}~\bibnamefont
  {Prada}}, \bibinfo {author} {\bibfnamefont {P.}~\bibnamefont {San-Jose}},
  \bibinfo {author} {\bibfnamefont {M.~W.~A.}\ \bibnamefont {de~Moor}},
  \bibinfo {author} {\bibfnamefont {A.}~\bibnamefont {Geresdi}}, \bibinfo
  {author} {\bibfnamefont {E.~J.~H.}\ \bibnamefont {Lee}}, \bibinfo {author}
  {\bibfnamefont {J.}~\bibnamefont {Klinovaja}}, \bibinfo {author}
  {\bibfnamefont {D.}~\bibnamefont {Loss}}, \bibinfo {author} {\bibfnamefont
  {J.}~\bibnamefont {Nyg{\aa}rd}}, \bibinfo {author} {\bibfnamefont
  {R.}~\bibnamefont {Aguado}},\ and\ \bibinfo {author} {\bibfnamefont {L.~P.}\
  \bibnamefont {Kouwenhoven}},\ }\bibfield  {title} {\bibinfo {title} {From
  {Andreev} to {M}ajorana bound states in hybrid superconductor--semiconductor
  nanowires},\ }\href {https://doi.org/10.1038/s42254-020-0228-y} {\bibfield
  {journal} {\bibinfo  {journal} {Nat. Rev. Phys.}\ }\textbf {\bibinfo {volume}
  {2}},\ \bibinfo {pages} {575} (\bibinfo {year} {2020})}\BibitemShut {NoStop}%
\bibitem [{\citenamefont {Laubscher}\ and\ \citenamefont
  {Klinovaja}(2021)}]{Laub2021}%
  \BibitemOpen
  \bibfield  {author} {\bibinfo {author} {\bibfnamefont {K.}~\bibnamefont
  {Laubscher}}\ and\ \bibinfo {author} {\bibfnamefont {J.}~\bibnamefont
  {Klinovaja}},\ }\bibfield  {title} {\bibinfo {title} {{M}ajorana bound states
  in semiconducting nanostructures},\ }\href
  {https://doi.org/10.1063/5.0055997} {\bibfield  {journal} {\bibinfo
  {journal} {J. Appl. Phys.}\ }\textbf {\bibinfo {volume} {130}},\ \bibinfo
  {pages} {081101} (\bibinfo {year} {2021})}\BibitemShut {NoStop}%
\bibitem [{\citenamefont {Zeng}\ \emph {et~al.}(2018)\citenamefont {Zeng},
  \citenamefont {Lei}, \citenamefont {Chaudhary},\ and\ \citenamefont
  {MacDonald}}]{Zeng18}%
  \BibitemOpen
  \bibfield  {author} {\bibinfo {author} {\bibfnamefont {Y.}~\bibnamefont
  {Zeng}}, \bibinfo {author} {\bibfnamefont {C.}~\bibnamefont {Lei}}, \bibinfo
  {author} {\bibfnamefont {G.}~\bibnamefont {Chaudhary}},\ and\ \bibinfo
  {author} {\bibfnamefont {A.~H.}\ \bibnamefont {MacDonald}},\ }\bibfield
  {title} {\bibinfo {title} {Quantum anomalous {H}all {M}ajorana platform},\
  }\href {https://doi.org/10.1103/PhysRevB.97.081102} {\bibfield  {journal}
  {\bibinfo  {journal} {Phys. Rev. B}\ }\textbf {\bibinfo {volume} {97}},\
  \bibinfo {pages} {081102} (\bibinfo {year} {2018})}\BibitemShut {NoStop}%
\bibitem [{\citenamefont {Chen}\ \emph {et~al.}(2018)\citenamefont {Chen},
  \citenamefont {Xie}, \citenamefont {Liu}, \citenamefont {Lee},\ and\
  \citenamefont {Law}}]{Chen18}%
  \BibitemOpen
  \bibfield  {author} {\bibinfo {author} {\bibfnamefont {C.-Z.}\ \bibnamefont
  {Chen}}, \bibinfo {author} {\bibfnamefont {Y.-M.}\ \bibnamefont {Xie}},
  \bibinfo {author} {\bibfnamefont {J.}~\bibnamefont {Liu}}, \bibinfo {author}
  {\bibfnamefont {P.~A.}\ \bibnamefont {Lee}},\ and\ \bibinfo {author}
  {\bibfnamefont {K.~T.}\ \bibnamefont {Law}},\ }\bibfield  {title} {\bibinfo
  {title} {Quasi-one-dimensional quantum anomalous {H}all systems as new
  platforms for scalable topological quantum computation},\ }\href
  {https://doi.org/10.1103/PhysRevB.97.104504} {\bibfield  {journal} {\bibinfo
  {journal} {Phys. Rev. B}\ }\textbf {\bibinfo {volume} {97}},\ \bibinfo
  {pages} {104504} (\bibinfo {year} {2018})}\BibitemShut {NoStop}%
\bibitem [{\citenamefont {Atanov}\ \emph {et~al.}(2024)\citenamefont {Atanov},
  \citenamefont {Tai}, \citenamefont {Xie}, \citenamefont {Ng}, \citenamefont
  {Hammond}, \citenamefont {{Manfred Ho}}, \citenamefont {Koo}, \citenamefont
  {Li}, \citenamefont {Ho}, \citenamefont {Lyu}, \citenamefont {Chong},
  \citenamefont {Zhang}, \citenamefont {Tai}, \citenamefont {Wang},
  \citenamefont {Law}, \citenamefont {Wang},\ and\ \citenamefont
  {Lortz}}]{Atan2023}%
  \BibitemOpen
  \bibfield  {author} {\bibinfo {author} {\bibfnamefont {O.}~\bibnamefont
  {Atanov}}, \bibinfo {author} {\bibfnamefont {W.~T.}\ \bibnamefont {Tai}},
  \bibinfo {author} {\bibfnamefont {Y.-M.}\ \bibnamefont {Xie}}, \bibinfo
  {author} {\bibfnamefont {Y.~H.}\ \bibnamefont {Ng}}, \bibinfo {author}
  {\bibfnamefont {M.~A.}\ \bibnamefont {Hammond}}, \bibinfo {author}
  {\bibfnamefont {T.~S.}\ \bibnamefont {{Manfred Ho}}}, \bibinfo {author}
  {\bibfnamefont {T.~H.}\ \bibnamefont {Koo}}, \bibinfo {author} {\bibfnamefont
  {H.}~\bibnamefont {Li}}, \bibinfo {author} {\bibfnamefont {S.~L.}\
  \bibnamefont {Ho}}, \bibinfo {author} {\bibfnamefont {J.}~\bibnamefont
  {Lyu}}, \bibinfo {author} {\bibfnamefont {S.}~\bibnamefont {Chong}}, \bibinfo
  {author} {\bibfnamefont {P.}~\bibnamefont {Zhang}}, \bibinfo {author}
  {\bibfnamefont {L.}~\bibnamefont {Tai}}, \bibinfo {author} {\bibfnamefont
  {J.}~\bibnamefont {Wang}}, \bibinfo {author} {\bibfnamefont {K.~T.}\
  \bibnamefont {Law}}, \bibinfo {author} {\bibfnamefont {K.~L.}\ \bibnamefont
  {Wang}},\ and\ \bibinfo {author} {\bibfnamefont {R.}~\bibnamefont {Lortz}},\
  }\bibfield  {title} {\bibinfo {title} {Proximity-induced
  quasi-one-dimensional superconducting quantum anomalous {H}all state},\
  }\href {https://doi.org/https://doi.org/10.1016/j.xcrp.2023.101762}
  {\bibfield  {journal} {\bibinfo  {journal} {Cell Rep. Phys. Sci.}\ }\textbf
  {\bibinfo {volume} {5}},\ \bibinfo {pages} {101762} (\bibinfo {year}
  {2024})}\BibitemShut {NoStop}%
\bibitem [{\citenamefont {Nadj-Perge}\ \emph {et~al.}(2014)\citenamefont
  {Nadj-Perge}, \citenamefont {Drozdov}, \citenamefont {Li}, \citenamefont
  {Chen}, \citenamefont {Jeon}, \citenamefont {Seo}, \citenamefont {MacDonald},
  \citenamefont {Bernevig},\ and\ \citenamefont {Yazdani}}]{Nadj2014}%
  \BibitemOpen
  \bibfield  {author} {\bibinfo {author} {\bibfnamefont {S.}~\bibnamefont
  {Nadj-Perge}}, \bibinfo {author} {\bibfnamefont {I.~K.}\ \bibnamefont
  {Drozdov}}, \bibinfo {author} {\bibfnamefont {J.}~\bibnamefont {Li}},
  \bibinfo {author} {\bibfnamefont {H.}~\bibnamefont {Chen}}, \bibinfo {author}
  {\bibfnamefont {S.}~\bibnamefont {Jeon}}, \bibinfo {author} {\bibfnamefont
  {J.}~\bibnamefont {Seo}}, \bibinfo {author} {\bibfnamefont {A.~H.}\
  \bibnamefont {MacDonald}}, \bibinfo {author} {\bibfnamefont {B.~A.}\
  \bibnamefont {Bernevig}},\ and\ \bibinfo {author} {\bibfnamefont
  {A.}~\bibnamefont {Yazdani}},\ }\bibfield  {title} {\bibinfo {title}
  {Observation of {M}ajorana fermions in ferromagnetic atomic chains on a
  superconductor},\ }\href {https://doi.org/10.1126/science.1259327} {\bibfield
   {journal} {\bibinfo  {journal} {Science}\ }\textbf {\bibinfo {volume}
  {346}},\ \bibinfo {pages} {602} (\bibinfo {year} {2014})}\BibitemShut
  {NoStop}%
\bibitem [{\citenamefont {Sau}\ and\ \citenamefont {Sarma}(2012)}]{Sau2012}%
  \BibitemOpen
  \bibfield  {author} {\bibinfo {author} {\bibfnamefont {J.~D.}\ \bibnamefont
  {Sau}}\ and\ \bibinfo {author} {\bibfnamefont {S.~D.}\ \bibnamefont
  {Sarma}},\ }\bibfield  {title} {\bibinfo {title} {Realizing a robust
  practical {M}ajorana chain in a quantum-dot-superconductor linear array},\
  }\href {https://doi.org/10.1038/ncomms1966} {\bibfield  {journal} {\bibinfo
  {journal} {Nat. Commun.}\ }\textbf {\bibinfo {volume} {3}},\ \bibinfo {pages}
  {964} (\bibinfo {year} {2012})}\BibitemShut {NoStop}%
\bibitem [{\citenamefont {Qi}\ \emph {et~al.}(2010)\citenamefont {Qi},
  \citenamefont {Hughes},\ and\ \citenamefont {Zhang}}]{Qi2010}%
  \BibitemOpen
  \bibfield  {author} {\bibinfo {author} {\bibfnamefont {X.-L.}\ \bibnamefont
  {Qi}}, \bibinfo {author} {\bibfnamefont {T.~L.}\ \bibnamefont {Hughes}},\
  and\ \bibinfo {author} {\bibfnamefont {S.-C.}\ \bibnamefont {Zhang}},\
  }\bibfield  {title} {\bibinfo {title} {Chiral topological superconductor from
  the quantum {H}all state},\ }\href
  {https://doi.org/10.1103/PhysRevB.82.184516} {\bibfield  {journal} {\bibinfo
  {journal} {Phys. Rev. B}\ }\textbf {\bibinfo {volume} {82}},\ \bibinfo
  {pages} {184516} (\bibinfo {year} {2010})}\BibitemShut {NoStop}%
\bibitem [{\citenamefont {K\"{o}nig}\ \emph {et~al.}(2008)\citenamefont
  {K\"{o}nig}, \citenamefont {Buhmann}, \citenamefont {W.~Molenkamp},
  \citenamefont {Hughes}, \citenamefont {Liu}, \citenamefont {Qi},\ and\
  \citenamefont {Zhang}}]{Kon2008}%
  \BibitemOpen
  \bibfield  {author} {\bibinfo {author} {\bibfnamefont {M.}~\bibnamefont
  {K\"{o}nig}}, \bibinfo {author} {\bibfnamefont {H.}~\bibnamefont {Buhmann}},
  \bibinfo {author} {\bibfnamefont {L.}~\bibnamefont {W.~Molenkamp}}, \bibinfo
  {author} {\bibfnamefont {T.}~\bibnamefont {Hughes}}, \bibinfo {author}
  {\bibfnamefont {C.-X.}\ \bibnamefont {Liu}}, \bibinfo {author} {\bibfnamefont
  {X.-L.}\ \bibnamefont {Qi}},\ and\ \bibinfo {author} {\bibfnamefont {S.-C.}\
  \bibnamefont {Zhang}},\ }\bibfield  {title} {\bibinfo {title} {The quantum
  spin {H}all effect: Theory and experiment},\ }\href
  {https://doi.org/10.1143/JPSJ.77.031007} {\bibfield  {journal} {\bibinfo
  {journal} {J. Phys. Soc. Jpn.}\ }\textbf {\bibinfo {volume} {77}},\ \bibinfo
  {pages} {031007} (\bibinfo {year} {2008})}\BibitemShut {NoStop}%
\bibitem [{\citenamefont {Chang}\ \emph {et~al.}(2013)\citenamefont {Chang},
  \citenamefont {Zhang}, \citenamefont {Feng}, \citenamefont {Shen},
  \citenamefont {Zhang}, \citenamefont {Guo}, \citenamefont {Li}, \citenamefont
  {Ou}, \citenamefont {Wei}, \citenamefont {Wang}, \citenamefont {Ji},
  \citenamefont {Feng}, \citenamefont {Ji}, \citenamefont {Chen}, \citenamefont
  {Jia}, \citenamefont {Dai}, \citenamefont {Fang}, \citenamefont {Zhang},
  \citenamefont {He}, \citenamefont {Wang}, \citenamefont {Lu}, \citenamefont
  {Ma},\ and\ \citenamefont {Xue}}]{Chang2013}%
  \BibitemOpen
  \bibfield  {author} {\bibinfo {author} {\bibfnamefont {C.-Z.}\ \bibnamefont
  {Chang}}, \bibinfo {author} {\bibfnamefont {J.}~\bibnamefont {Zhang}},
  \bibinfo {author} {\bibfnamefont {X.}~\bibnamefont {Feng}}, \bibinfo {author}
  {\bibfnamefont {J.}~\bibnamefont {Shen}}, \bibinfo {author} {\bibfnamefont
  {Z.}~\bibnamefont {Zhang}}, \bibinfo {author} {\bibfnamefont
  {M.}~\bibnamefont {Guo}}, \bibinfo {author} {\bibfnamefont {K.}~\bibnamefont
  {Li}}, \bibinfo {author} {\bibfnamefont {Y.}~\bibnamefont {Ou}}, \bibinfo
  {author} {\bibfnamefont {P.}~\bibnamefont {Wei}}, \bibinfo {author}
  {\bibfnamefont {L.-L.}\ \bibnamefont {Wang}}, \bibinfo {author}
  {\bibfnamefont {Z.-Q.}\ \bibnamefont {Ji}}, \bibinfo {author} {\bibfnamefont
  {Y.}~\bibnamefont {Feng}}, \bibinfo {author} {\bibfnamefont {S.}~\bibnamefont
  {Ji}}, \bibinfo {author} {\bibfnamefont {X.}~\bibnamefont {Chen}}, \bibinfo
  {author} {\bibfnamefont {J.}~\bibnamefont {Jia}}, \bibinfo {author}
  {\bibfnamefont {X.}~\bibnamefont {Dai}}, \bibinfo {author} {\bibfnamefont
  {Z.}~\bibnamefont {Fang}}, \bibinfo {author} {\bibfnamefont {S.-C.}\
  \bibnamefont {Zhang}}, \bibinfo {author} {\bibfnamefont {K.}~\bibnamefont
  {He}}, \bibinfo {author} {\bibfnamefont {Y.}~\bibnamefont {Wang}}, \bibinfo
  {author} {\bibfnamefont {L.}~\bibnamefont {Lu}}, \bibinfo {author}
  {\bibfnamefont {X.-C.}\ \bibnamefont {Ma}},\ and\ \bibinfo {author}
  {\bibfnamefont {Q.-K.}\ \bibnamefont {Xue}},\ }\bibfield  {title} {\bibinfo
  {title} {Experimental observation of the quantum anomalous {H}all effect in a
  magnetic topological insulator},\ }\href
  {https://doi.org/10.1126/science.1234414} {\bibfield  {journal} {\bibinfo
  {journal} {Science}\ }\textbf {\bibinfo {volume} {340}},\ \bibinfo {pages}
  {167} (\bibinfo {year} {2013})}\BibitemShut {NoStop}%
\bibitem [{\citenamefont {Qiu}\ \emph {et~al.}(2022)\citenamefont {Qiu},
  \citenamefont {Zhang}, \citenamefont {Deng}, \citenamefont {Chong},
  \citenamefont {Tai}, \citenamefont {Eckberg},\ and\ \citenamefont
  {Wang}}]{Qiu2022}%
  \BibitemOpen
  \bibfield  {author} {\bibinfo {author} {\bibfnamefont {G.}~\bibnamefont
  {Qiu}}, \bibinfo {author} {\bibfnamefont {P.}~\bibnamefont {Zhang}}, \bibinfo
  {author} {\bibfnamefont {P.}~\bibnamefont {Deng}}, \bibinfo {author}
  {\bibfnamefont {S.~K.}\ \bibnamefont {Chong}}, \bibinfo {author}
  {\bibfnamefont {L.}~\bibnamefont {Tai}}, \bibinfo {author} {\bibfnamefont
  {C.}~\bibnamefont {Eckberg}},\ and\ \bibinfo {author} {\bibfnamefont {K.~L.}\
  \bibnamefont {Wang}},\ }\bibfield  {title} {\bibinfo {title} {Mesoscopic
  transport of quantum anomalous {H}all effect in the submicron size regime},\
  }\href {https://doi.org/10.1103/PhysRevLett.128.217704} {\bibfield  {journal}
  {\bibinfo  {journal} {Phys. Rev. Lett.}\ }\textbf {\bibinfo {volume} {128}},\
  \bibinfo {pages} {217704} (\bibinfo {year} {2022})}\BibitemShut {NoStop}%
\bibitem [{\citenamefont {Nayak}\ \emph {et~al.}(2008)\citenamefont {Nayak},
  \citenamefont {Simon}, \citenamefont {Stern}, \citenamefont {Freedman},\ and\
  \citenamefont {Das~Sarma}}]{Nay08}%
  \BibitemOpen
  \bibfield  {author} {\bibinfo {author} {\bibfnamefont {C.}~\bibnamefont
  {Nayak}}, \bibinfo {author} {\bibfnamefont {S.~H.}\ \bibnamefont {Simon}},
  \bibinfo {author} {\bibfnamefont {A.}~\bibnamefont {Stern}}, \bibinfo
  {author} {\bibfnamefont {M.}~\bibnamefont {Freedman}},\ and\ \bibinfo
  {author} {\bibfnamefont {S.}~\bibnamefont {Das~Sarma}},\ }\bibfield  {title}
  {\bibinfo {title} {Non-abelian anyons and topological quantum computation},\
  }\href {https://doi.org/10.1103/RevModPhys.80.1083} {\bibfield  {journal}
  {\bibinfo  {journal} {Rev. Mod. Phys.}\ }\textbf {\bibinfo {volume} {80}},\
  \bibinfo {pages} {1083} (\bibinfo {year} {2008})}\BibitemShut {NoStop}%
\bibitem [{\citenamefont {Kayyalha}\ \emph {et~al.}(2020)\citenamefont
  {Kayyalha}, \citenamefont {Xiao}, \citenamefont {Zhang}, \citenamefont
  {Shin}, \citenamefont {Jiang}, \citenamefont {Wang}, \citenamefont {Zhao},
  \citenamefont {Xiao}, \citenamefont {Zhang}, \citenamefont {Fijalkowski},
  \citenamefont {Mandal}, \citenamefont {Winnerlein}, \citenamefont {Gould},
  \citenamefont {Li}, \citenamefont {Molenkamp}, \citenamefont {Chan},
  \citenamefont {Samarth},\ and\ \citenamefont {Chang}}]{Kay2020}%
  \BibitemOpen
  \bibfield  {author} {\bibinfo {author} {\bibfnamefont {M.}~\bibnamefont
  {Kayyalha}}, \bibinfo {author} {\bibfnamefont {D.}~\bibnamefont {Xiao}},
  \bibinfo {author} {\bibfnamefont {R.}~\bibnamefont {Zhang}}, \bibinfo
  {author} {\bibfnamefont {J.}~\bibnamefont {Shin}}, \bibinfo {author}
  {\bibfnamefont {J.}~\bibnamefont {Jiang}}, \bibinfo {author} {\bibfnamefont
  {F.}~\bibnamefont {Wang}}, \bibinfo {author} {\bibfnamefont {Y.-F.}\
  \bibnamefont {Zhao}}, \bibinfo {author} {\bibfnamefont {R.}~\bibnamefont
  {Xiao}}, \bibinfo {author} {\bibfnamefont {L.}~\bibnamefont {Zhang}},
  \bibinfo {author} {\bibfnamefont {K.~M.}\ \bibnamefont {Fijalkowski}},
  \bibinfo {author} {\bibfnamefont {P.}~\bibnamefont {Mandal}}, \bibinfo
  {author} {\bibfnamefont {M.}~\bibnamefont {Winnerlein}}, \bibinfo {author}
  {\bibfnamefont {C.}~\bibnamefont {Gould}}, \bibinfo {author} {\bibfnamefont
  {Q.}~\bibnamefont {Li}}, \bibinfo {author} {\bibfnamefont {L.~W.}\
  \bibnamefont {Molenkamp}}, \bibinfo {author} {\bibfnamefont {M.~H.~W.}\
  \bibnamefont {Chan}}, \bibinfo {author} {\bibfnamefont {N.}~\bibnamefont
  {Samarth}},\ and\ \bibinfo {author} {\bibfnamefont {C.-Z.}\ \bibnamefont
  {Chang}},\ }\bibfield  {title} {\bibinfo {title} {Absence of evidence for
  chiral {M}ajorana modes in quantum anomalous {H}all-superconductor devices},\
  }\href {https://doi.org/10.1126/science.aax6361} {\bibfield  {journal}
  {\bibinfo  {journal} {Science}\ }\textbf {\bibinfo {volume} {367}},\ \bibinfo
  {pages} {64} (\bibinfo {year} {2020})}\BibitemShut {NoStop}%
\bibitem [{\citenamefont {Quantum}(2023)}]{Agh2023}%
  \BibitemOpen
  \bibfield  {author} {\bibinfo {author} {\bibfnamefont {M.}~\bibnamefont
  {Quantum}},\ }\bibfield  {title} {\bibinfo {title} {{I}n{A}s-{A}l hybrid
  devices passing the topological gap protocol},\ }\href
  {https://doi.org/10.1103/PhysRevB.107.245423} {\bibfield  {journal} {\bibinfo
   {journal} {Phys. Rev. B}\ }\textbf {\bibinfo {volume} {107}},\ \bibinfo
  {pages} {245423} (\bibinfo {year} {2023})}\BibitemShut {NoStop}%
\bibitem [{\citenamefont {Uday}\ \emph {et~al.}(2024)\citenamefont {Uday},
  \citenamefont {Lippertz}, \citenamefont {Moors}, \citenamefont {Legg},
  \citenamefont {Joris}, \citenamefont {Bliesener}, \citenamefont {Pereira},
  \citenamefont {Taskin},\ and\ \citenamefont {Ando}}]{Uday2024}%
  \BibitemOpen
  \bibfield  {author} {\bibinfo {author} {\bibfnamefont {A.}~\bibnamefont
  {Uday}}, \bibinfo {author} {\bibfnamefont {G.}~\bibnamefont {Lippertz}},
  \bibinfo {author} {\bibfnamefont {K.}~\bibnamefont {Moors}}, \bibinfo
  {author} {\bibfnamefont {H.~F.}\ \bibnamefont {Legg}}, \bibinfo {author}
  {\bibfnamefont {R.}~\bibnamefont {Joris}}, \bibinfo {author} {\bibfnamefont
  {A.}~\bibnamefont {Bliesener}}, \bibinfo {author} {\bibfnamefont {L.~M.~C.}\
  \bibnamefont {Pereira}}, \bibinfo {author} {\bibfnamefont {A.~A.}\
  \bibnamefont {Taskin}},\ and\ \bibinfo {author} {\bibfnamefont
  {Y.}~\bibnamefont {Ando}},\ }\bibfield  {title} {\bibinfo {title} {Induced
  superconducting correlations in a quantum anomalous {H}all insulator},\
  }\href {https://doi.org/10.1038/s41567-024-02574-1} {\bibfield  {journal}
  {\bibinfo  {journal} {Nat. Phys.}\ }\textbf {\bibinfo {volume} {20}},\
  \bibinfo {pages} {1589} (\bibinfo {year} {2024})}\BibitemShut {NoStop}%
\bibitem [{\citenamefont {Uday}\ \emph {et~al.}(2025)\citenamefont {Uday},
  \citenamefont {Lippertz}, \citenamefont {Bhujel}, \citenamefont {Taskin},\
  and\ \citenamefont {Ando}}]{Uday2025}%
  \BibitemOpen
  \bibfield  {author} {\bibinfo {author} {\bibfnamefont {A.}~\bibnamefont
  {Uday}}, \bibinfo {author} {\bibfnamefont {G.}~\bibnamefont {Lippertz}},
  \bibinfo {author} {\bibfnamefont {B.}~\bibnamefont {Bhujel}}, \bibinfo
  {author} {\bibfnamefont {A.~A.}\ \bibnamefont {Taskin}},\ and\ \bibinfo
  {author} {\bibfnamefont {Y.}~\bibnamefont {Ando}},\ }\bibfield  {title}
  {\bibinfo {title} {Non-{M}ajorana origin of the half-integer conductance
  quantization elucidated by multiterminal superconductor--quantum anomalous
  {H}all insulator heterostructure},\ }\href
  {https://doi.org/10.1103/PhysRevB.111.035440} {\bibfield  {journal} {\bibinfo
   {journal} {Phys. Rev. B}\ }\textbf {\bibinfo {volume} {111}},\ \bibinfo
  {pages} {035440} (\bibinfo {year} {2025})}\BibitemShut {NoStop}%
\bibitem [{\citenamefont {Quantum}(2025)}]{Agh2025}%
  \BibitemOpen
  \bibfield  {author} {\bibinfo {author} {\bibfnamefont {M.~A.}\ \bibnamefont
  {Quantum}},\ }\bibfield  {title} {\bibinfo {title} {Interferometric
  single-shot parity measurement in {I}n{A}s--{A}l hybrid devices},\ }\href
  {https://doi.org/10.1038/s41586-024-08445-2} {\bibfield  {journal} {\bibinfo
  {journal} {Nature}\ }\textbf {\bibinfo {volume} {638}},\ \bibinfo {pages}
  {651} (\bibinfo {year} {2025})}\BibitemShut {NoStop}%
\bibitem [{\citenamefont {Tokura}\ \emph {et~al.}(2019)\citenamefont {Tokura},
  \citenamefont {Yasuda},\ and\ \citenamefont {Tsukazaki}}]{Toku2019}%
  \BibitemOpen
  \bibfield  {author} {\bibinfo {author} {\bibfnamefont {Y.}~\bibnamefont
  {Tokura}}, \bibinfo {author} {\bibfnamefont {K.}~\bibnamefont {Yasuda}},\
  and\ \bibinfo {author} {\bibfnamefont {A.}~\bibnamefont {Tsukazaki}},\
  }\bibfield  {title} {\bibinfo {title} {Magnetic topological insulators},\
  }\href {https://doi.org/10.1038/s42254-018-0011-5} {\bibfield  {journal}
  {\bibinfo  {journal} {Nat. Rev. Phys.}\ }\textbf {\bibinfo {volume} {1}},\
  \bibinfo {pages} {126} (\bibinfo {year} {2019})}\BibitemShut {NoStop}%
\bibitem [{\citenamefont {Chang}\ \emph {et~al.}(2023)\citenamefont {Chang},
  \citenamefont {Liu},\ and\ \citenamefont {MacDonald}}]{Chang2023}%
  \BibitemOpen
  \bibfield  {author} {\bibinfo {author} {\bibfnamefont {C.-Z.}\ \bibnamefont
  {Chang}}, \bibinfo {author} {\bibfnamefont {C.-X.}\ \bibnamefont {Liu}},\
  and\ \bibinfo {author} {\bibfnamefont {A.~H.}\ \bibnamefont {MacDonald}},\
  }\bibfield  {title} {\bibinfo {title} {Colloquium: Quantum anomalous {H}all
  effect},\ }\href {https://doi.org/10.1103/RevModPhys.95.011002} {\bibfield
  {journal} {\bibinfo  {journal} {Rev. Mod. Phys.}\ }\textbf {\bibinfo {volume}
  {95}},\ \bibinfo {pages} {011002} (\bibinfo {year} {2023})}\BibitemShut
  {NoStop}%
\bibitem [{\citenamefont {Fu}\ and\ \citenamefont {Kane}(2008)}]{Fu2008}%
  \BibitemOpen
  \bibfield  {author} {\bibinfo {author} {\bibfnamefont {L.}~\bibnamefont
  {Fu}}\ and\ \bibinfo {author} {\bibfnamefont {C.~L.}\ \bibnamefont {Kane}},\
  }\bibfield  {title} {\bibinfo {title} {Superconducting proximity effect and
  {{M}ajorana} fermions at the surface of a topological insulator},\ }\href
  {https://doi.org/10.1103/PhysRevLett.100.096407} {\bibfield  {journal}
  {\bibinfo  {journal} {Phys. Rev. Lett.}\ }\textbf {\bibinfo {volume} {100}},\
  \bibinfo {pages} {096407} (\bibinfo {year} {2008})}\BibitemShut {NoStop}%
\bibitem [{\citenamefont {Wang}\ \emph
  {et~al.}(2015{\natexlab{a}})\citenamefont {Wang}, \citenamefont {Zhou},
  \citenamefont {Lian},\ and\ \citenamefont {Zhang}}]{Wang2015}%
  \BibitemOpen
  \bibfield  {author} {\bibinfo {author} {\bibfnamefont {J.}~\bibnamefont
  {Wang}}, \bibinfo {author} {\bibfnamefont {Q.}~\bibnamefont {Zhou}}, \bibinfo
  {author} {\bibfnamefont {B.}~\bibnamefont {Lian}},\ and\ \bibinfo {author}
  {\bibfnamefont {S.-C.}\ \bibnamefont {Zhang}},\ }\bibfield  {title} {\bibinfo
  {title} {Chiral topological superconductor and half-integer conductance
  plateau from quantum anomalous {H}all plateau transition},\ }\href
  {https://doi.org/10.1103/PhysRevB.92.064520} {\bibfield  {journal} {\bibinfo
  {journal} {Phys. Rev. B}\ }\textbf {\bibinfo {volume} {92}},\ \bibinfo
  {pages} {064520} (\bibinfo {year} {2015}{\natexlab{a}})}\BibitemShut
  {NoStop}%
\bibitem [{\citenamefont {Osca}\ and\ \citenamefont {Serra}(2018)}]{Osca2018}%
  \BibitemOpen
  \bibfield  {author} {\bibinfo {author} {\bibfnamefont {J.}~\bibnamefont
  {Osca}}\ and\ \bibinfo {author} {\bibfnamefont {L.}~\bibnamefont {Serra}},\
  }\bibfield  {title} {\bibinfo {title} {Conductance oscillations and speed of
  chiral {M}ajorana mode in a quantum-anomalous-{H}all 2d strip},\ }\href
  {https://doi.org/10.1103/PhysRevB.98.121407} {\bibfield  {journal} {\bibinfo
  {journal} {Phys. Rev. B}\ }\textbf {\bibinfo {volume} {98}},\ \bibinfo
  {pages} {121407} (\bibinfo {year} {2018})}\BibitemShut {NoStop}%
\bibitem [{\citenamefont {Wang}\ \emph
  {et~al.}(2015{\natexlab{b}})\citenamefont {Wang}, \citenamefont {Lian},\ and\
  \citenamefont {Zhang}}]{Wang2015b}%
  \BibitemOpen
  \bibfield  {author} {\bibinfo {author} {\bibfnamefont {J.}~\bibnamefont
  {Wang}}, \bibinfo {author} {\bibfnamefont {B.}~\bibnamefont {Lian}},\ and\
  \bibinfo {author} {\bibfnamefont {S.-C.}\ \bibnamefont {Zhang}},\ }\bibfield
  {title} {\bibinfo {title} {Electrically tunable magnetism in magnetic
  topological insulators},\ }\href
  {https://doi.org/10.1103/PhysRevLett.115.036805} {\bibfield  {journal}
  {\bibinfo  {journal} {Phys. Rev. Lett.}\ }\textbf {\bibinfo {volume} {115}},\
  \bibinfo {pages} {036805} (\bibinfo {year} {2015}{\natexlab{b}})}\BibitemShut
  {NoStop}%
\bibitem [{\citenamefont {Wang}(2016)}]{Wang2016}%
  \BibitemOpen
  \bibfield  {author} {\bibinfo {author} {\bibfnamefont {J.}~\bibnamefont
  {Wang}},\ }\bibfield  {title} {\bibinfo {title} {Electrically tunable
  topological superconductivity and {M}ajorana fermions in two dimensions},\
  }\href {https://doi.org/10.1103/PhysRevB.94.214502} {\bibfield  {journal}
  {\bibinfo  {journal} {Phys. Rev. B}\ }\textbf {\bibinfo {volume} {94}},\
  \bibinfo {pages} {214502} (\bibinfo {year} {2016})}\BibitemShut {NoStop}%
\bibitem [{\citenamefont {Chong}\ \emph {et~al.}(2023)\citenamefont {Chong},
  \citenamefont {Zhang}, \citenamefont {Li}, \citenamefont {Zhou},
  \citenamefont {Wang}, \citenamefont {Zhang}, \citenamefont {Davydov},
  \citenamefont {Eckberg}, \citenamefont {Deng}, \citenamefont {Tai},
  \citenamefont {Xia}, \citenamefont {Wu},\ and\ \citenamefont
  {Wang}}]{Chong2023}%
  \BibitemOpen
  \bibfield  {author} {\bibinfo {author} {\bibfnamefont {S.~K.}\ \bibnamefont
  {Chong}}, \bibinfo {author} {\bibfnamefont {P.}~\bibnamefont {Zhang}},
  \bibinfo {author} {\bibfnamefont {J.}~\bibnamefont {Li}}, \bibinfo {author}
  {\bibfnamefont {Y.}~\bibnamefont {Zhou}}, \bibinfo {author} {\bibfnamefont
  {J.}~\bibnamefont {Wang}}, \bibinfo {author} {\bibfnamefont {H.}~\bibnamefont
  {Zhang}}, \bibinfo {author} {\bibfnamefont {A.~V.}\ \bibnamefont {Davydov}},
  \bibinfo {author} {\bibfnamefont {C.}~\bibnamefont {Eckberg}}, \bibinfo
  {author} {\bibfnamefont {P.}~\bibnamefont {Deng}}, \bibinfo {author}
  {\bibfnamefont {L.}~\bibnamefont {Tai}}, \bibinfo {author} {\bibfnamefont
  {J.}~\bibnamefont {Xia}}, \bibinfo {author} {\bibfnamefont {R.}~\bibnamefont
  {Wu}},\ and\ \bibinfo {author} {\bibfnamefont {K.~L.}\ \bibnamefont {Wang}},\
  }\bibfield  {title} {\bibinfo {title} {Electrical manipulation of topological
  phases in a quantum anomalous {H}all insulator},\ }\href
  {https://doi.org/https://doi.org/10.1002/adma.202207622} {\bibfield
  {journal} {\bibinfo  {journal} {Adv. Mater.}\ }\textbf {\bibinfo {volume}
  {35}},\ \bibinfo {pages} {2207622} (\bibinfo {year} {2023})}\BibitemShut
  {NoStop}%
\bibitem [{\citenamefont {Wang}\ \emph {et~al.}(2013)\citenamefont {Wang},
  \citenamefont {Lian}, \citenamefont {Zhang},\ and\ \citenamefont
  {Zhang}}]{Wang2013}%
  \BibitemOpen
  \bibfield  {author} {\bibinfo {author} {\bibfnamefont {J.}~\bibnamefont
  {Wang}}, \bibinfo {author} {\bibfnamefont {B.}~\bibnamefont {Lian}}, \bibinfo
  {author} {\bibfnamefont {H.}~\bibnamefont {Zhang}},\ and\ \bibinfo {author}
  {\bibfnamefont {S.-C.}\ \bibnamefont {Zhang}},\ }\bibfield  {title} {\bibinfo
  {title} {Anomalous edge transport in the quantum anomalous {H}all state},\
  }\href {https://doi.org/10.1103/PhysRevLett.111.086803} {\bibfield  {journal}
  {\bibinfo  {journal} {Phys. Rev. Lett.}\ }\textbf {\bibinfo {volume} {111}},\
  \bibinfo {pages} {086803} (\bibinfo {year} {2013})}\BibitemShut {NoStop}%
\bibitem [{\citenamefont {Wang}\ \emph {et~al.}(2014)\citenamefont {Wang},
  \citenamefont {Lian},\ and\ \citenamefont {Zhang}}]{Wang2014}%
  \BibitemOpen
  \bibfield  {author} {\bibinfo {author} {\bibfnamefont {J.}~\bibnamefont
  {Wang}}, \bibinfo {author} {\bibfnamefont {B.}~\bibnamefont {Lian}},\ and\
  \bibinfo {author} {\bibfnamefont {S.-C.}\ \bibnamefont {Zhang}},\ }\bibfield
  {title} {\bibinfo {title} {Universal scaling of the quantum anomalous {H}all
  plateau transition},\ }\href {https://doi.org/10.1103/PhysRevB.89.085106}
  {\bibfield  {journal} {\bibinfo  {journal} {Phys. Rev. B}\ }\textbf {\bibinfo
  {volume} {89}},\ \bibinfo {pages} {085106} (\bibinfo {year}
  {2014})}\BibitemShut {NoStop}%
\bibitem [{\citenamefont {Lian}\ \emph {et~al.}(2016)\citenamefont {Lian},
  \citenamefont {Wang},\ and\ \citenamefont {Zhang}}]{Lian2016}%
  \BibitemOpen
  \bibfield  {author} {\bibinfo {author} {\bibfnamefont {B.}~\bibnamefont
  {Lian}}, \bibinfo {author} {\bibfnamefont {J.}~\bibnamefont {Wang}},\ and\
  \bibinfo {author} {\bibfnamefont {S.-C.}\ \bibnamefont {Zhang}},\ }\bibfield
  {title} {\bibinfo {title} {Edge-state-induced {Andreev} oscillation in
  quantum anomalous {H}all insulator-superconductor junctions},\ }\href
  {https://doi.org/10.1103/PhysRevB.93.161401} {\bibfield  {journal} {\bibinfo
  {journal} {Phys. Rev. B}\ }\textbf {\bibinfo {volume} {93}},\ \bibinfo
  {pages} {161401} (\bibinfo {year} {2016})}\BibitemShut {NoStop}%
\bibitem [{\citenamefont {Legendre}\ \emph {et~al.}(2024)\citenamefont
  {Legendre}, \citenamefont {Zsurka}, \citenamefont {Di~Miceli}, \citenamefont
  {Serra}, \citenamefont {Moors},\ and\ \citenamefont {Schmidt}}]{Legend2024}%
  \BibitemOpen
  \bibfield  {author} {\bibinfo {author} {\bibfnamefont {J.}~\bibnamefont
  {Legendre}}, \bibinfo {author} {\bibfnamefont {E.}~\bibnamefont {Zsurka}},
  \bibinfo {author} {\bibfnamefont {D.}~\bibnamefont {Di~Miceli}}, \bibinfo
  {author} {\bibfnamefont {L.}~\bibnamefont {Serra}}, \bibinfo {author}
  {\bibfnamefont {K.}~\bibnamefont {Moors}},\ and\ \bibinfo {author}
  {\bibfnamefont {T.~L.}\ \bibnamefont {Schmidt}},\ }\bibfield  {title}
  {\bibinfo {title} {Topological properties of finite-size heterostructures of
  magnetic topological insulators and superconductors},\ }\href
  {https://doi.org/10.1103/PhysRevB.110.075426} {\bibfield  {journal} {\bibinfo
   {journal} {Phys. Rev. B}\ }\textbf {\bibinfo {volume} {110}},\ \bibinfo
  {pages} {075426} (\bibinfo {year} {2024})}\BibitemShut {NoStop}%
\bibitem [{\citenamefont {Zsurka}\ \emph {et~al.}(2024)\citenamefont {Zsurka},
  \citenamefont {Wang}, \citenamefont {Legendre}, \citenamefont {Di~Miceli},
  \citenamefont {Serra}, \citenamefont {Gr\"utzmacher}, \citenamefont
  {Schmidt}, \citenamefont {R\"u\ss{}mann},\ and\ \citenamefont
  {Moors}}]{Zsurka2024}%
  \BibitemOpen
  \bibfield  {author} {\bibinfo {author} {\bibfnamefont {E.}~\bibnamefont
  {Zsurka}}, \bibinfo {author} {\bibfnamefont {C.}~\bibnamefont {Wang}},
  \bibinfo {author} {\bibfnamefont {J.}~\bibnamefont {Legendre}}, \bibinfo
  {author} {\bibfnamefont {D.}~\bibnamefont {Di~Miceli}}, \bibinfo {author}
  {\bibfnamefont {L.}~\bibnamefont {Serra}}, \bibinfo {author} {\bibfnamefont
  {D.}~\bibnamefont {Gr\"utzmacher}}, \bibinfo {author} {\bibfnamefont {T.~L.}\
  \bibnamefont {Schmidt}}, \bibinfo {author} {\bibfnamefont {P.}~\bibnamefont
  {R\"u\ss{}mann}},\ and\ \bibinfo {author} {\bibfnamefont {K.}~\bibnamefont
  {Moors}},\ }\bibfield  {title} {\bibinfo {title} {Low-energy modeling of
  three-dimensional topological insulator nanostructures},\ }\href
  {https://doi.org/10.1103/PhysRevMaterials.8.084204} {\bibfield  {journal}
  {\bibinfo  {journal} {Phys. Rev. Mater.}\ }\textbf {\bibinfo {volume} {8}},\
  \bibinfo {pages} {084204} (\bibinfo {year} {2024})}\BibitemShut {NoStop}%
\bibitem [{\citenamefont {Osca}\ and\ \citenamefont {Serra}(2019)}]{Osca2019}%
  \BibitemOpen
  \bibfield  {author} {\bibinfo {author} {\bibfnamefont {J.}~\bibnamefont
  {Osca}}\ and\ \bibinfo {author} {\bibfnamefont {L.}~\bibnamefont {Serra}},\
  }\bibfield  {title} {\bibinfo {title} {Complex band-structure analysis and
  topological physics of {M}ajorana nanowires},\ }\href
  {https://doi.org/10.1140/epjb/e2019-100011-2} {\bibfield  {journal} {\bibinfo
   {journal} {Eur. Phys. J. B}\ }\textbf {\bibinfo {volume} {92}},\ \bibinfo
  {pages} {101} (\bibinfo {year} {2019})}\BibitemShut {NoStop}%
\bibitem [{\citenamefont {Di~Miceli}\ \emph {et~al.}(2023)\citenamefont
  {Di~Miceli}, \citenamefont {Zsurka}, \citenamefont {Legendre}, \citenamefont
  {Moors}, \citenamefont {Schmidt},\ and\ \citenamefont {Serra}}]{DiMi2023}%
  \BibitemOpen
  \bibfield  {author} {\bibinfo {author} {\bibfnamefont {D.}~\bibnamefont
  {Di~Miceli}}, \bibinfo {author} {\bibfnamefont {E.}~\bibnamefont {Zsurka}},
  \bibinfo {author} {\bibfnamefont {J.}~\bibnamefont {Legendre}}, \bibinfo
  {author} {\bibfnamefont {K.}~\bibnamefont {Moors}}, \bibinfo {author}
  {\bibfnamefont {T.~L.}\ \bibnamefont {Schmidt}},\ and\ \bibinfo {author}
  {\bibfnamefont {L.}~\bibnamefont {Serra}},\ }\bibfield  {title} {\bibinfo
  {title} {Conductance asymmetry in proximitized magnetic topological insulator
  junctions with {M}ajorana modes},\ }\href
  {https://doi.org/10.1103/PhysRevB.108.035424} {\bibfield  {journal} {\bibinfo
   {journal} {Phys. Rev. B}\ }\textbf {\bibinfo {volume} {108}},\ \bibinfo
  {pages} {035424} (\bibinfo {year} {2023})}\BibitemShut {NoStop}%
\bibitem [{\citenamefont {Lambert}\ \emph {et~al.}(1993)\citenamefont
  {Lambert}, \citenamefont {Hui},\ and\ \citenamefont {Robinson}}]{Lamb1993}%
  \BibitemOpen
  \bibfield  {author} {\bibinfo {author} {\bibfnamefont {C.}~\bibnamefont
  {Lambert}}, \bibinfo {author} {\bibfnamefont {V.}~\bibnamefont {Hui}},\ and\
  \bibinfo {author} {\bibfnamefont {S.}~\bibnamefont {Robinson}},\ }\bibfield
  {title} {\bibinfo {title} {Multi-probe conductance formulae for mesoscopic
  superconductors},\ }\href {https://doi.org/10.1088/0953-8984/5/25/009}
  {\bibfield  {journal} {\bibinfo  {journal} {J. Phys.: Condens. Matter}\
  }\textbf {\bibinfo {volume} {5}},\ \bibinfo {pages} {4187} (\bibinfo {year}
  {1993})}\BibitemShut {NoStop}%
\bibitem [{\citenamefont {Nadeem}\ \emph {et~al.}(2023)\citenamefont {Nadeem},
  \citenamefont {Fuhrer},\ and\ \citenamefont {Wang}}]{Nade2023}%
  \BibitemOpen
  \bibfield  {author} {\bibinfo {author} {\bibfnamefont {M.}~\bibnamefont
  {Nadeem}}, \bibinfo {author} {\bibfnamefont {M.~S.}\ \bibnamefont {Fuhrer}},\
  and\ \bibinfo {author} {\bibfnamefont {X.}~\bibnamefont {Wang}},\ }\bibfield
  {title} {\bibinfo {title} {The superconducting diode effect},\ }\href
  {https://doi.org/10.1038/s42254-023-00632-w} {\bibfield  {journal} {\bibinfo
  {journal} {Nat. Rev. Phys.}\ }\textbf {\bibinfo {volume} {5}},\ \bibinfo
  {pages} {558} (\bibinfo {year} {2023})}\BibitemShut {NoStop}%
\bibitem [{\citenamefont {Ji}\ and\ \citenamefont {Wen}(2018)}]{Ji2018}%
  \BibitemOpen
  \bibfield  {author} {\bibinfo {author} {\bibfnamefont {W.}~\bibnamefont
  {Ji}}\ and\ \bibinfo {author} {\bibfnamefont {X.-G.}\ \bibnamefont {Wen}},\
  }\bibfield  {title} {\bibinfo {title} {$\frac{1}{2}({e}^{2}/h)$ conductance
  plateau without 1d chiral {M}ajorana fermions},\ }\href
  {https://doi.org/10.1103/PhysRevLett.120.107002} {\bibfield  {journal}
  {\bibinfo  {journal} {Phys. Rev. Lett.}\ }\textbf {\bibinfo {volume} {120}},\
  \bibinfo {pages} {107002} (\bibinfo {year} {2018})}\BibitemShut {NoStop}%
\bibitem [{\citenamefont {Huang}\ \emph {et~al.}(2018)\citenamefont {Huang},
  \citenamefont {Setiawan},\ and\ \citenamefont {Sau}}]{Huang2018}%
  \BibitemOpen
  \bibfield  {author} {\bibinfo {author} {\bibfnamefont {Y.}~\bibnamefont
  {Huang}}, \bibinfo {author} {\bibfnamefont {F.}~\bibnamefont {Setiawan}},\
  and\ \bibinfo {author} {\bibfnamefont {J.~D.}\ \bibnamefont {Sau}},\
  }\bibfield  {title} {\bibinfo {title} {Disorder-induced half-integer
  quantized conductance plateau in quantum anomalous {H}all
  insulator-superconductor structures},\ }\href
  {https://doi.org/10.1103/PhysRevB.97.100501} {\bibfield  {journal} {\bibinfo
  {journal} {Phys. Rev. B}\ }\textbf {\bibinfo {volume} {97}},\ \bibinfo
  {pages} {100501} (\bibinfo {year} {2018})}\BibitemShut {NoStop}%
\bibitem [{\citenamefont {Lian}\ \emph {et~al.}(2018)\citenamefont {Lian},
  \citenamefont {Sun}, \citenamefont {Vaezi}, \citenamefont {Qi},\ and\
  \citenamefont {Zhang}}]{Lian2018}%
  \BibitemOpen
  \bibfield  {author} {\bibinfo {author} {\bibfnamefont {B.}~\bibnamefont
  {Lian}}, \bibinfo {author} {\bibfnamefont {X.-Q.}\ \bibnamefont {Sun}},
  \bibinfo {author} {\bibfnamefont {A.}~\bibnamefont {Vaezi}}, \bibinfo
  {author} {\bibfnamefont {X.-L.}\ \bibnamefont {Qi}},\ and\ \bibinfo {author}
  {\bibfnamefont {S.-C.}\ \bibnamefont {Zhang}},\ }\bibfield  {title} {\bibinfo
  {title} {Topological quantum computation based on chiral {M}ajorana
  fermions},\ }\href {https://doi.org/10.1073/pnas.1810003115} {\bibfield
  {journal} {\bibinfo  {journal} {Proc. Natl. Acad. Sci.}\ }\textbf {\bibinfo
  {volume} {115}},\ \bibinfo {pages} {10938} (\bibinfo {year}
  {2018})}\BibitemShut {NoStop}%
\bibitem [{\citenamefont {Di~Miceli}\ and\ \citenamefont
  {Serra}(2023)}]{Dimi2023b}%
  \BibitemOpen
  \bibfield  {author} {\bibinfo {author} {\bibfnamefont {D.}~\bibnamefont
  {Di~Miceli}}\ and\ \bibinfo {author} {\bibfnamefont {L.}~\bibnamefont
  {Serra}},\ }\bibfield  {title} {\bibinfo {title} {Quantum-anomalous-{H}all
  current patterns and interference in thin slabs of chiral topological
  superconductors},\ }\href {https://doi.org/10.1038/s41598-023-47286-3}
  {\bibfield  {journal} {\bibinfo  {journal} {Sci. Rep.}\ }\textbf {\bibinfo
  {volume} {13}},\ \bibinfo {pages} {19955} (\bibinfo {year}
  {2023})}\BibitemShut {NoStop}%
\end{thebibliography}%

\clearpage
\appendix
\section{Algorithm details}
\label{appA}

We have solved the scattering problem mentioned in Sec.\ \ref{scatt}
using the complex band structure 
formalism \cite{Osca2019}. 
This approach is well suited
for describing piecewise homogenous systems, such as the MTI slab with 
several regions of Fig.\ \ref{F0}. The method is computationally 
efficient as it does not require spatial
2D ($xy$) grids, but rather a 1D transverse grid ($y$) and a set of 
complex $k$ eigenmodes for each region of uniform parameters. 
Consequently, the computational demand is independent of the lengths
$L_g$, $\ell$ and $L_S$ of the regions forming the junction.
The algorithm is a two-step process. First, the complex $k$'s and their corresponding eigenmodes are obtained for each region by diagonalizing an eigenvalue problem derived from the transformation of the 
Bogoliubov-deGennes $E$-eigenproblem
\begin{equation}
\label{eqA1}
{\cal H}_k \Psi_k(y) = E \Psi_k(y)\;, 
\end{equation}
where $H_k$ is the Hamiltonian of Eq.\ (\ref{eq1}) with the replacement
$p_x\to\hbar k$. 
The transformed $k$-eigenproblem reads \cite{DiMi2023}
\begin{eqnarray}
\left(
\begin{array}{cc}
0 & 1 \\
-{\cal C}^{-1}({\cal A}-E) &
-{\cal C}^{-1}\cal B
\end{array}
\right)
\,
\left(
\begin{array}{c}
\Phi_k(y)\\
k\,\Phi_k(y)
\end{array}
\right)
&=& \nonumber\\
&&
\!\!\!\!\!\!\!\!\!\!\!\!
{\hspace*{-2cm}}
k\,
\left(
\begin{array}{c}
\Phi_k(y)\\
k\,\Phi_k(y)
\end{array}
\right)
\label{eqA2}
\; ,
\end{eqnarray}
In Eq.\ (\ref{eqA2}) we have grouped the different contributions to the Hamiltonian,
Eq.\ (\ref{eq1}), by defining the
operators ${\cal A}$, ${\cal B}$ and ${\cal C}$ as follows:
\begin{eqnarray}
{\cal A} &=&
\left(\, m_0 + m_1 p_y^2\, \right) \lambda_x\,\tau_z  
\nonumber\\
&+& \frac{\alpha}{\hbar}\, p_y\sigma_x\, 
\lambda_z \,\tau_z
+ \Delta_Z\, \sigma_z
-\mu\,\tau_z
\nonumber\\
&+& \left( \Delta_p + 
\Delta_m\,\lambda_z\right) \tau_x\,
+\Delta_g\,\lambda_z\,\tau_z\; ,\\
{\cal B} &=& -\alpha\, \sigma_y\lambda_z\tau_z\;,\\
 {\cal C} &=& m_1\, \hbar^2 \lambda_x\tau_z\; .
 \end{eqnarray}

Notice that Eq.\ (\ref{eqA2}) is an eigenvalue problem in an enlarged space by a factor of two, due to the additional eigenvector component $k\Phi_k$.
Furthermore, since the operator of Eq.\ (\ref{eqA2}) is non-Hermitian it yields $k$ eigenvalues 
that can have an imaginary part. Physically, these complex-$k$ eigenmodes are relevant close to the junction interfaces.

The second step of the algorithm involves determining
the set of amplitudes $\{ C_k^{i} \}$, 
see Eq.\ (\ref{eq2}), corresponding to the output modes in L and R
leads for a given input. In practice this involves solving a linear 
system obtained from the continuity conditions at the junction 
interfaces. The wave function must be continuous at the interfaces
along $x$. Since the Hamiltonian  Eq.\ (\ref{eq2}) contains a 
term in $p_x^2$, the 
first $x$-derivative should also be continuous at the interfaces. 
However, the small value of parameter $m_1$ means that the continuity condition 
on the first $x$-derivative is only a very small correction.

The number of $k$-eigenmodes in each region is truncated, including the $N_k$ modes having $|k|\le K$, where $K$ is the cutoff value. We typically include $N_k\approx 100$.
The continuity conditions at the interfaces are projected onto the set
of complex modes in order to yield a closed linear system for the 
set of unknowns $\{C_k\}$. The stability and convergence of the method
is controlled by increasing 
the number $N_k$ of complex modes and the number $N_y$ of $y$ grid 
points. 
The quasiparticle velocity operator $v_{qp}$ along the slab and the
associated flux $I_{qp}$
are defined as
\begin{eqnarray}
\label{eqA6}
v_{qp}& \equiv & \frac{\partial {\cal H}}{\partial p_x}
=
-\alpha\, \sigma_y\lambda_z\tau_z
+2 m_1\, p_x \lambda_x\tau_z\; , \\
\label{eqA7}
I_{qp} &=&
\Re{\int{dy\, \Psi^*(x,y)\, v_{qp}\, \Psi(x,y)}}
\; ,
\end{eqnarray}
where $\Re$ takes the real part.

The corresponding charge velocity operator $v_c$ and current $I_c$ 
are defined similarly to Eq.\ (\ref{eqA6}) and (\ref{eqA7}) 
adding an extra $\tau_z$ operator
in Eq.\ (\ref{eqA6}). The quasiparticle flux $I_{qp}$ 
should be conserved
but this is not necessarily true for the
charge flux $I_{c}$ due to the possible Andreev
electron-hole transformations.
As a general criterion, we make sure that quasiparticle flux conservation 
between input and output
is fulfilled 
by the algorithm with a precision better than 1\%.

\begin{figure*}[t]
\centering
\includegraphics[width=0.65\textwidth,trim=0.5cm 11cm 0.5cm 1cm,clip]{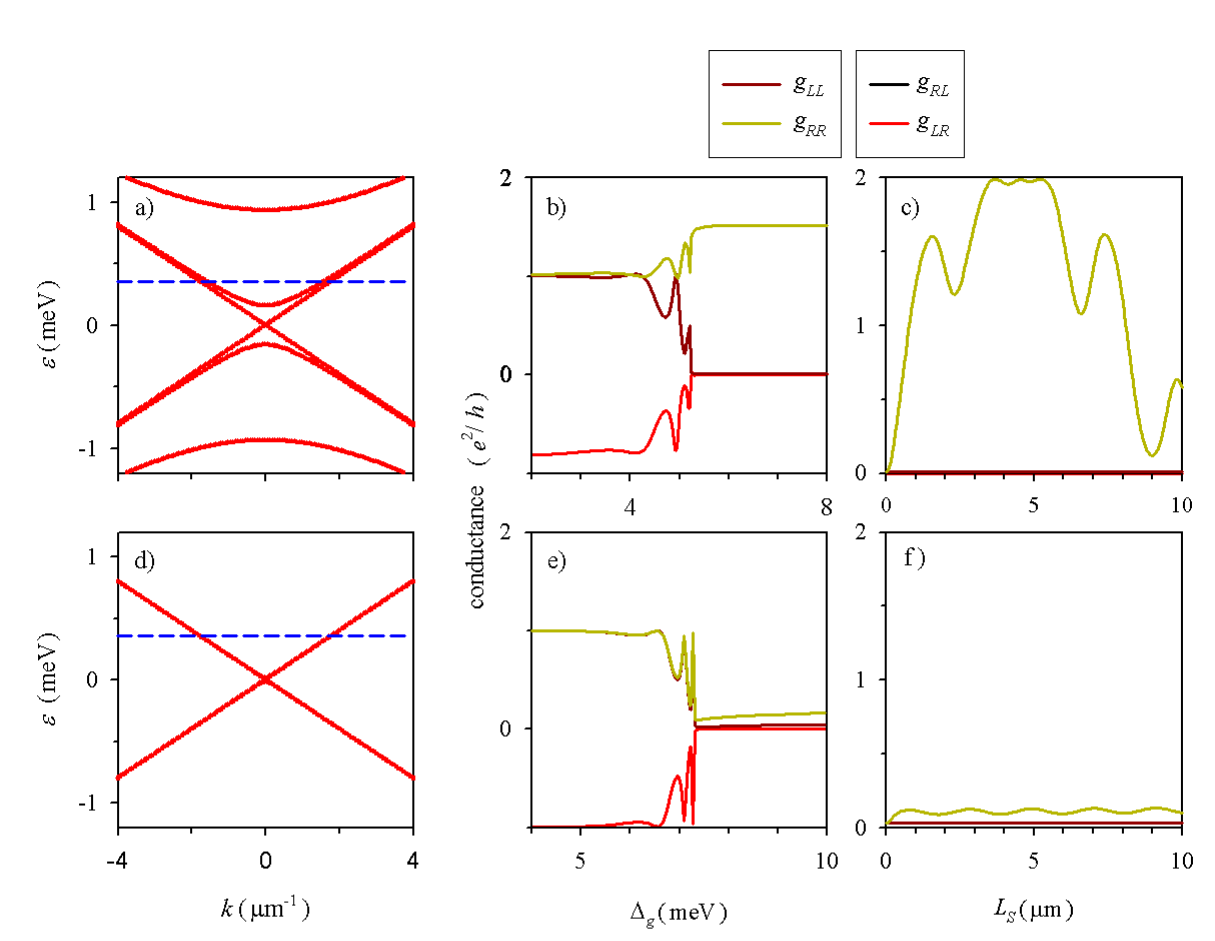}  
 \caption{
a,d) Band structure in the superconducting region for a slab of $L_y=1\, \mu{\rm m}$; 
b,c,e,f) Conductances for $L_g=7\, \mu{\rm m}$, $\ell=0$ as a function 
of $\Delta_g$ for a fixed $L_S=2.8\, \mu{\rm m}$ (b,e), and as a function of 
$L_S$ for a fixed $\Delta_g=8\,{\rm meV}$ (c,f).
The upper row of panels (a,b,c) is  with a magnetization 
$\Delta_Z=17.8\, {\rm meV}$, slightly below the ${\cal N}=2$ phase transition.
The lower row (d,e,f) is with $\Delta_Z=18.5\, {\rm meV}$, which is above the 
phase transition.
The dashed blue lines in panels a) and d) indicate the assumed transport energy 
$E=0.3\, {\rm meV}$.  
Note that the Dirac cross in panel d) is doubly degenerate.
The pairing strengths are $\Delta_1=1\; {\rm meV}$ and $\Delta_2=0$ in all cases.
Panels c) and f) 
are in the electrical cutoff regime due to the large $\Delta_g$. 
The interference oscillation with $L_S$ shows a multiple frequency beating  pattern in panel c).
This pattern 
indicates the activation of multiple modes, in contrast to
the case of a single Majorana mode.}
  \label{F10}
\end{figure*}

\section{${\cal N}=2$ physics}
\label{appB}

The precise ${\cal N}=2$ topological phase 
of a superconducting slab
corresponds to the emergence  
of a doubly degenerate Majorana cross in the $\varepsilon(k)$ spectrum 
of states. These two degenerate Majoranas are equivalent 
to a pure Fermionic mode. The transport to normal leads mediated by a 
Fermionic mode and that mediated by a Majorana mode showing stark differences \cite{Wang2015}. 
This sharp phase-transition scenario is of course modified by finite-size effects 
in a real slab with $L_y$ in the micron or submicron range. Here, the emergence 
of a pure Fermionic mode is more gradual as the magnetization increases,
transitioning
from a single Majorana mode regime to a two-Majorana mode regime.
An intermediate regime exists in which the two Majoranas are not exactly degenerate, resulting in conductance beating patterns \cite{Dimi2023b}. 

Figure\ \ref{F10} shows the band diagrams and conductances for two values 
of the slab magnetization $\Delta_Z$.
Both values are close to the phase transition point between ${\cal N}=1$ and ${\cal N}=2$,
where an additional pair of Majorana modes emerges at the device edges.
Deeper in the ${\cal N}=2$ phase the expected transport signature 
for large $L_y$
is equivalent to that of a 
QAH state in a slab without superconductivity. Therefore, in the
electrical cutoff regime we expect all conductances 
will vanish in our system. That is, the interference oscillation of $g_{RR}$ with $L_S$ typical 
of the  Majorana double reflection and interference ($M_{1,2}$ 
of Fig.\ \ref{F0}ed) will be washed out deep in the ${\cal N}=2$ phase.

As shown in Fig.\ \ref{F10},
slightly before the phase transition (upper panels) the Majorana bands are not 
degenerate, the local conductances
$g_{LL}$ and $g_{RR}$ differ and display correlated oscillations. In
particular, $g_{RR}$ exhibits a pattern of multiple-frequency beating as a function of $L_S$ in the 
cutoff regime (Fig.\ \ref{F10}c).
Conversely, when the magnetization is high enough, as shown in the lower panels of Fig.\ {\ref{F10}}, we recover 
the scenario of an electronic QAH state throughout the slab, 
with  identical 
$g_{LL}$ and $g_{RR}$ as a function of $\Delta_g$,
not exceeding one quantum of conductance (Fig.\ \ref{F10}e).
In the cutoff regime
$g_{RR}$ becomes small, washing out the oscillations with $L_S$ (Fig.\ \ref{F10}f). 
Increasing the magnetization further,
the degeneracy of the bands and the vanishing of conductances 
in the cutoff regime
are improved.

\end{document}